\documentclass{aastex63}

\usepackage{textcomp}
\usepackage[utf8]{inputenc}
\usepackage[titletoc, title]{appendix}
\usepackage{CJKutf8}
\usepackage{float}
\usepackage{blindtext}
\usepackage{gensymb}
\usepackage{xcolor}
\usepackage{hyperref}

\received{17 October 2020}
\revised{7 January 2021}
\accepted{9 February 2021}
\submitjournal{The Planetary Science Journal}

\def \Physics {Department of Physics, University of Michigan, Ann Arbor, MI 48109, USA}
\def \Astronomy {Department of Astronomy, University of Michigan, Ann Arbor, MI 48109, USA}
\def \NSFGRF {NSF Graduate Research Fellow}
\def \UPenn {Department of Physics and Astronomy, University of Pennsylvania, Philadelphia, PA 19104, USA}

\graphicspath{{Figures/}}

\begin{document}
\shorttitle{No Evidence for Clustering in the Extreme TNOs}
\shortauthors{Napier et al.}

\title{No Evidence for Orbital Clustering in the Extreme Trans-Neptunian Objects}

\correspondingauthor{Kevin J. Napier}
\email{kjnapier@umich.edu}

\author[0000-0003-4827-5049]{K.~J.~Napier}
\affiliation{\Physics}
\author[0000-0001-6942-2736]{D.~W.~Gerdes}
\affiliation{\Physics}
\affiliation{\Astronomy}
\author[0000-0001-7737-6784]{Hsing~Wen~Lin (\begin{CJK*}{UTF8}{gbsn}
林省文\end{CJK*})}
\affiliation{\Physics}
\author[0000-0002-6126-8487]{S.~J.~Hamilton}
\altaffiliation{\NSFGRF}
\affiliation{\Physics}
\author[0000-0002-8613-8259]{G.~M.~Bernstein}
\affiliation{\UPenn}
\author[0000-0003-0743-9422]{P.~H.~Bernardinelli}
\affiliation{\UPenn}

\author{T.~M.~C.~Abbott}
\affiliation{Cerro Tololo Inter-American Observatory, NSF's National Optical-Infrared Astronomy Research Laboratory, Casilla 603, La Serena, Chile}
\author{M.~Aguena}
\affiliation{Departamento de F\'isica Matem\'atica, Instituto de F\'isica, Universidade de S\~ao Paulo, CP 66318, S\~ao Paulo, SP, 05314-970, Brazil}
\affiliation{Laborat\'orio Interinstitucional de e-Astronomia - LIneA, Rua Gal. Jos\'e Cristino 77, Rio de Janeiro, RJ - 20921-400, Brazil}
\author{J.~Annis}
\affiliation{Fermi National Accelerator Laboratory, P. O. Box 500, Batavia, IL 60510, USA}
\author{S.~Avila}
\affiliation{Instituto de Fisica Teorica UAM/CSIC, Universidad Autonoma de Madrid, 28049 Madrid, Spain}
\author{D.~Bacon}
\affiliation{Institute of Cosmology and Gravitation, University of Portsmouth, Portsmouth, PO1 3FX, UK}
\author{E.~Bertin}
\affiliation{CNRS, UMR 7095, Institut d'Astrophysique de Paris, F-75014, Paris, France}
\affiliation{Sorbonne Universit\'es, UPMC Univ Paris 06, UMR 7095, Institut d'Astrophysique de Paris, F-75014, Paris, France}
\author{D.~Brooks}
\affiliation{Department of Physics \& Astronomy, University College London, Gower Street, London, WC1E 6BT, UK}
\author{D.~L.~Burke}
\affiliation{Kavli Institute for Particle Astrophysics \& Cosmology, P. O. Box 2450, Stanford University, Stanford, CA 94305, USA}
\affiliation{SLAC National Accelerator Laboratory, Menlo Park, CA 94025, USA}
\author{A.~Carnero~Rosell}
\affiliation{Instituto de Astrofisica de Canarias, E-38205 La Laguna, Tenerife, Spain}
\affiliation{Universidad de La Laguna, Dpto. Astrofísica, E-38206 La Laguna, Tenerife, Spain}
\author{M.~Carrasco~Kind}
\affiliation{Department of Astronomy, University of Illinois at Urbana-Champaign, 1002 W. Green Street, Urbana, IL 61801, USA}
\affiliation{National Center for Supercomputing Applications, 1205 West Clark St., Urbana, IL 61801, USA}
\author{J.~Carretero}
\affiliation{Institut de F\'{\i}sica d'Altes Energies (IFAE), The Barcelona Institute of Science and Technology, Campus UAB, 08193 Bellaterra (Barcelona) Spain}
\author{M.~Costanzi}
\affiliation{INAF-Osservatorio Astronomico di Trieste, via G. B. Tiepolo 11, I-34143 Trieste, Italy}
\affiliation{Institute for Fundamental Physics of the Universe, Via Beirut 2, 34014 Trieste, Italy}
\author{L.~N.~da Costa}
\affiliation{Laborat\'orio Interinstitucional de e-Astronomia - LIneA, Rua Gal. Jos\'e Cristino 77, Rio de Janeiro, RJ - 20921-400, Brazil}
\affiliation{Observat\'orio Nacional, Rua Gal. Jos\'e Cristino 77, Rio de Janeiro, RJ - 20921-400, Brazil}
\author{J.~De~Vicente}
\affiliation{Centro de Investigaciones Energ\'eticas, Medioambientales y Tecnol\'ogicas (CIEMAT), Madrid, Spain}
\author{H.~T.~Diehl}
\affiliation{Fermi National Accelerator Laboratory, P. O. Box 500, Batavia, IL 60510, USA}
\author{P.~Doel}
\affiliation{Department of Physics \& Astronomy, University College London, Gower Street, London, WC1E 6BT, UK}
\author{S.~Everett}
\affiliation{Santa Cruz Institute for Particle Physics, Santa Cruz, CA 95064, USA}
\author{I.~Ferrero}
\affiliation{Institute of Theoretical Astrophysics, University of Oslo. P.O. Box 1029 Blindern, NO-0315 Oslo, Norway}
\author{P.~Fosalba}
\affiliation{Institut d'Estudis Espacials de Catalunya (IEEC), 08034 Barcelona, Spain}
\affiliation{Institute of Space Sciences (ICE, CSIC),  Campus UAB, Carrer de Can Magrans, s/n,  08193 Barcelona, Spain}
\author{J.~Garc\'ia-Bellido}
\affiliation{Instituto de Fisica Teorica UAM/CSIC, Universidad Autonoma de Madrid, 28049 Madrid, Spain}
\author{D.~Gruen}
\affiliation{Department of Physics, Stanford University, 382 Via Pueblo Mall, Stanford, CA 94305, USA}
\affiliation{Kavli Institute for Particle Astrophysics \& Cosmology, P. O. Box 2450, Stanford University, Stanford, CA 94305, USA}
\affiliation{SLAC National Accelerator Laboratory, Menlo Park, CA 94025, USA}
\author{R.~A.~Gruendl}
\affiliation{Department of Astronomy, University of Illinois at Urbana-Champaign, 1002 W. Green Street, Urbana, IL 61801, USA}
\affiliation{National Center for Supercomputing Applications, 1205 West Clark St., Urbana, IL 61801, USA}
\author{G.~Gutierrez}
\affiliation{Fermi National Accelerator Laboratory, P. O. Box 500, Batavia, IL 60510, USA}
\author{D.~L.~Hollowood}
\affiliation{Santa Cruz Institute for Particle Physics, Santa Cruz, CA 95064, USA}
\author{K.~Honscheid}
\affiliation{Center for Cosmology and Astro-Particle Physics, The Ohio State University, Columbus, OH 43210, USA}
\affiliation{Department of Physics, The Ohio State University, Columbus, OH 43210, USA}
\author{B.~Hoyle}
\affiliation{Faculty of Physics, Ludwig-Maximilians-Universit\"at, Scheinerstr. 1, 81679 Munich, Germany}
\affiliation{Max Planck Institute for Extraterrestrial Physics, Giessenbachstrasse, 85748 Garching, Germany}
\affiliation{Universit\"ats-Sternwarte, Fakult\"at f\"ur Physik, Ludwig-Maximilians Universit\"at M\"unchen, Scheinerstr. 1, 81679 M\"unchen, Germany}
\author{D.~J.~James}
\affiliation{Center for Astrophysics $\vert$ Harvard \& Smithsonian, 60 Garden Street, Cambridge, MA 02138, USA}
\author{S.~Kent}
\affiliation{Fermi National Accelerator Laboratory, P. O. Box 500, Batavia, IL 60510, USA}
\affiliation{Kavli Institute for Cosmological Physics, University of Chicago, Chicago, IL 60637, USA}
\author{K.~Kuehn}
\affiliation{Australian Astronomical Optics, Macquarie University, North Ryde, NSW 2113, Australia}
\affiliation{Lowell Observatory, 1400 Mars Hill Rd, Flagstaff, AZ 86001, USA}
\author{N.~Kuropatkin}
\affiliation{Fermi National Accelerator Laboratory, P. O. Box 500, Batavia, IL 60510, USA}
\author{M.~A.~G.~Maia}
\affiliation{Laborat\'orio Interinstitucional de e-Astronomia - LIneA, Rua Gal. Jos\'e Cristino 77, Rio de Janeiro, RJ - 20921-400, Brazil}
\affiliation{Observat\'orio Nacional, Rua Gal. Jos\'e Cristino 77, Rio de Janeiro, RJ - 20921-400, Brazil}
\author{F.~Menanteau}
\affiliation{Department of Astronomy, University of Illinois at Urbana-Champaign, 1002 W. Green Street, Urbana, IL 61801, USA}
\affiliation{National Center for Supercomputing Applications, 1205 West Clark St., Urbana, IL 61801, USA}
\author{R.~Miquel}
\affiliation{Instituci\'o Catalana de Recerca i Estudis Avan\c{c}ats, E-08010 Barcelona, Spain}
\affiliation{Institut de F\'{\i}sica d'Altes Energies (IFAE), The Barcelona Institute of Science and Technology, Campus UAB, 08193 Bellaterra (Barcelona) Spain}
\author{R.~Morgan}
\affiliation{Physics Department, 2320 Chamberlin Hall, University of Wisconsin-Madison, 1150 University Avenue Madison, WI  53706-1390}
\author{A.~Palmese}
\affiliation{Fermi National Accelerator Laboratory, P. O. Box 500, Batavia, IL 60510, USA}
\affiliation{Kavli Institute for Cosmological Physics, University of Chicago, Chicago, IL 60637, USA}
\author{F.~Paz-Chinch\'{o}n}
\affiliation{Institute of Astronomy, University of Cambridge, Madingley Road, Cambridge CB3 0HA, UK}
\affiliation{National Center for Supercomputing Applications, 1205 West Clark St., Urbana, IL 61801, USA}
\author{A.~A.~Plazas}
\affiliation{Department of Astrophysical Sciences, Princeton University, Peyton Hall, Princeton, NJ 08544, USA}
\author{E.~Sanchez}
\affiliation{Centro de Investigaciones Energ\'eticas, Medioambientales y Tecnol\'ogicas (CIEMAT), Madrid, Spain}
\author{V.~Scarpine}
\affiliation{Fermi National Accelerator Laboratory, P. O. Box 500, Batavia, IL 60510, USA}
\author{S.~Serrano}
\affiliation{Institut d'Estudis Espacials de Catalunya (IEEC), 08034 Barcelona, Spain}
\affiliation{Institute of Space Sciences (ICE, CSIC),  Campus UAB, Carrer de Can Magrans, s/n,  08193 Barcelona, Spain}
\author{I.~Sevilla-Noarbe}
\affiliation{Centro de Investigaciones Energ\'eticas, Medioambientales y Tecnol\'ogicas (CIEMAT), Madrid, Spain}
\author{M.~Smith}
\affiliation{School of Physics and Astronomy, University of Southampton,  Southampton, SO17 1BJ, UK}
\author{E.~Suchyta}
\affiliation{Computer Science and Mathematics Division, Oak Ridge National Laboratory, Oak Ridge, TN 37831}
\author{M.~E.~C.~Swanson}
\affiliation{National Center for Supercomputing Applications, 1205 West Clark St., Urbana, IL 61801, USA}
\author{C.~To}
\affiliation{Department of Physics, Stanford University, 382 Via Pueblo Mall, Stanford, CA 94305, USA}
\affiliation{Kavli Institute for Particle Astrophysics \& Cosmology, P. O. Box 2450, Stanford University, Stanford, CA 94305, USA}
\affiliation{SLAC National Accelerator Laboratory, Menlo Park, CA 94025, USA}
\author{A.~R.~Walker}
\affiliation{Cerro Tololo Inter-American Observatory, NSF's National Optical-Infrared Astronomy Research Laboratory, Casilla 603, La Serena, Chile}
\author{R.D.~Wilkinson}
\affiliation{Department of Physics and Astronomy, Pevensey Building, University of Sussex, Brighton, BN1 9QH, UK}

\collaboration{53}{(DES Collaboration)}

\begin{abstract}

The apparent clustering in longitude of perihelion $\varpi$ and ascending node $\Omega$ of extreme trans-Neptunian objects (ETNOs) has been attributed to the gravitational effects of an unseen 5-10 Earth-mass planet in the outer solar system. To investigate how selection bias may contribute to this clustering, we consider 14 ETNOs discovered by the Dark Energy Survey, the Outer Solar System Origins Survey, and the survey of Sheppard and Trujillo. Using each survey's published pointing history, depth, and TNO tracking selections, we calculate the joint probability that these objects are consistent with an underlying parent population with uniform distributions in $\varpi$ and $\Omega$. We find that the mean scaled longitude of perihelion and orbital poles of the detected ETNOs are consistent with a uniform population at a level between $17\%$ and $94\%$, and thus conclude that this sample provides no evidence for angular clustering.
\end{abstract}


\section{Introduction}
\label{sec:intro}

The apparent clustering in longitude of perihelion $\varpi$ and ascending node $\Omega$ of solar system bodies known as extreme trans-Neptunian objects (ETNOs) motivated the hypothesis that the solar system contains a 5-10 Earth-mass planet (Planet X/Planet 9) at 400-800 times Earth's distance from the sun \citep{st14, BB16, P9Review}. Some have proposed even more exotic sources of the apparent clustering, such as gravitational perturbations from a primordial black hole captured into orbit around the sun \citep{PhysRevLett.125.051103}.

While there is no universally accepted definition for the ETNOs, recent literature has emphasized objects with semi-major axis $a \gtrsim 230$ au and perihelion $q > 30$ au. Because ETNOs follow highly elliptical orbits, and their brightness decreases like $1/r^4$, they are almost always discovered within a few decades of perihelion. Moreover, telescopic surveys observe a limited area of the sky, at particular times of year, to a limited depth. These effects result in significant selection bias. The six ETNOs considered in the \citet{BB16} (BB16) analysis were discovered in an assortment of surveys with unknown or unpublished selection functions, making it difficult to establish that the observed angular clustering was indeed of physical origin.  

More recent surveys have carefully characterized their selection functions and applied these tools to small samples of new ETNOs. The Outer Solar System Origins Survey (OSSOS, \cite{Bannister2016}) analyzed the bias present in the discovery of 8 objects they detected with $a > 150$ au and $q > 30$ au \citep{OSSOS17}. They found that their detected objects were consistent with a uniform underlying population in $\varpi$ and $\Omega$. \citet{Bernardinelli2020} analyzed samples of 3-7 variously-defined ETNOs discovered by the Dark Energy Survey (DES, \cite{DES2005, 2016MNRAS.460.1270D}), and also found the data consistent with angular isotropy.

\citet{BB19} (BB19) attempted to reverse-engineer the survey bias in the entire then-known population of 14 ETNOs using a sampling method \citep{Brown2017} on all TNOs known to the Minor Planet Center. In contrast to the individual survey-level analyses described above, BB19 concluded that the observed clustering is highly likely to be a physical effect, and they argued that the best explanation remains a massive distant planet.

While no single survey has discovered enough ETNOs to reach a statistically compelling conclusion, a stronger statement becomes possible when data from multiple surveys are combined. According to the criteria above, there are 14 ETNOs (Table~\ref{tab:ETNOs}) detected by three independent surveys with characterized selection functions, all published since BB16. Using the published pointing history, depth, and TNO tracking selections for DES (5 objects) \citep{TaliClassification, BernardinelliCatalog}, OSSOS (5 objects) \citep{2018ApJS..236...18B}, and the survey of \citet{ST16} (ST) (4 objects), we calculate the joint probability that these objects are consistent with the null hypothesis: an underlying population distributed uniformly in the longitudes $\varpi$ and $\Omega$. If the purported clustering is indeed a physical effect, we would expect it to remain consistent with the data in this larger, independent sample when selection functions are modeled.

\begin{table}
\renewcommand{\arraystretch}{0.95}
\begin{tabular*}{\textwidth}{l@{\extracolsep{\fill}}rcrcrrcl}
\hline
\hline
Object & $a$ (au) & $e$ & $i$ ($\degree$) & $q$ (au) & $\omega$($\degree$) & $\Omega$ ($\degree$) & H (mag) & Survey \\ 
\hline
2015 BP$_{519}$ &  448.8   &  0.92  &  54.1 &  35.2 & 348.1 & 135.2 & 4.4 & DES \\
2013 SL$_{102}$ &  314.3   &  0.88  &  6.5  &  38.1 & 265.5 & 94.7  & 7.1 & DES \\
2013 RA$_{109}$ &  462.4   &  0.90  &  12.4 &  46.0 & 263.0 & 104.8 & 6.2 & DES \\
2014 WB$_{556}$ &  289.1   &  0.85  &  24.2 &  42.5 & 234.6 & 114.9 & 7.3 & DES \\
2016 SG$_{58}$  &  233.0   &  0.85  &  13.2 &  35.1 & 296.3 & 119.0 & 7.5 & DES \\
2013 SY$_{99}$  &  733.1   &  0.93  &  4.2  &  50.1 &  32.2 &  29.5 & 6.7 & OSSOS \\
2015 RX$_{245}$ &  426.4   &  0.89  &  12.1 &  45.7 &  65.1 &   8.6 & 6.2 & OSSOS \\
2015 GT$_{50}$  &  311.4   &  0.88  &  8.8  &  38.5 & 129.0 &  46.1 & 8.5 & OSSOS \\
2015 KG$_{163}$ &  679.7   &  0.94  &  14.0 &  40.5 &  32.1 & 219.1 & 8.2 & OSSOS \\
uo5m93          &  283.0   &  0.86  &  6.8  &  39.5 &  43.3 & 165.9 & 8.8 & OSSOS \\
2013 FT$_{28}$  &  295.4   &  0.85  &  17.4 &  43.4 &  40.7 & 217.7 & 6.7 & ST \\
2014 SR$_{349}$ &  296.6   &  0.84  &  18.0 &  47.7 & 341.2 &  34.9 & 6.7 & ST \\
2015 TG$_{387}$ & 1101.3   &  0.94  &  11.7 &  65.1  & 118.0 & 301.0 & 5.5 & ST \\
2014 FE$_{72}$  & 1559.5   &  0.98  &  20.6 &  36.2 & 133.9 & 336.8 & 6.1 & ST \\
2012 VP$_{113}$*  & 262.7   &  0.69  &  24.1 &  80.5 & 293.8 & 90.8 & 4.0 & ST \\
2013 RF$_{98}$* &  363.6   &  0.90  &  29.5 &  36.1 & 311.8 & 67.6  & 8.7 & DES SN \\
\hline
\end{tabular*}
\caption{\label{tab:ETNOs} Barycentric orbital elements of the ETNOs used in our analysis. All reported values are at the epoch JD $2459000.5$ (except for uo5m93 whose elements are for the epoch JD $2457163.82647$). DES SN indicates discovery in the DES supernova fields. *Note that in order to maintain an independent sample from BB16 we do not include 2013 RF$_{98}$ or 2012 VP$_{113}$ in our main analysis. We discuss their effects separately.}
\end{table}

\section{Methods}

\begin{figure}[ht]
    \centering
    \includegraphics[width=0.65\textwidth]{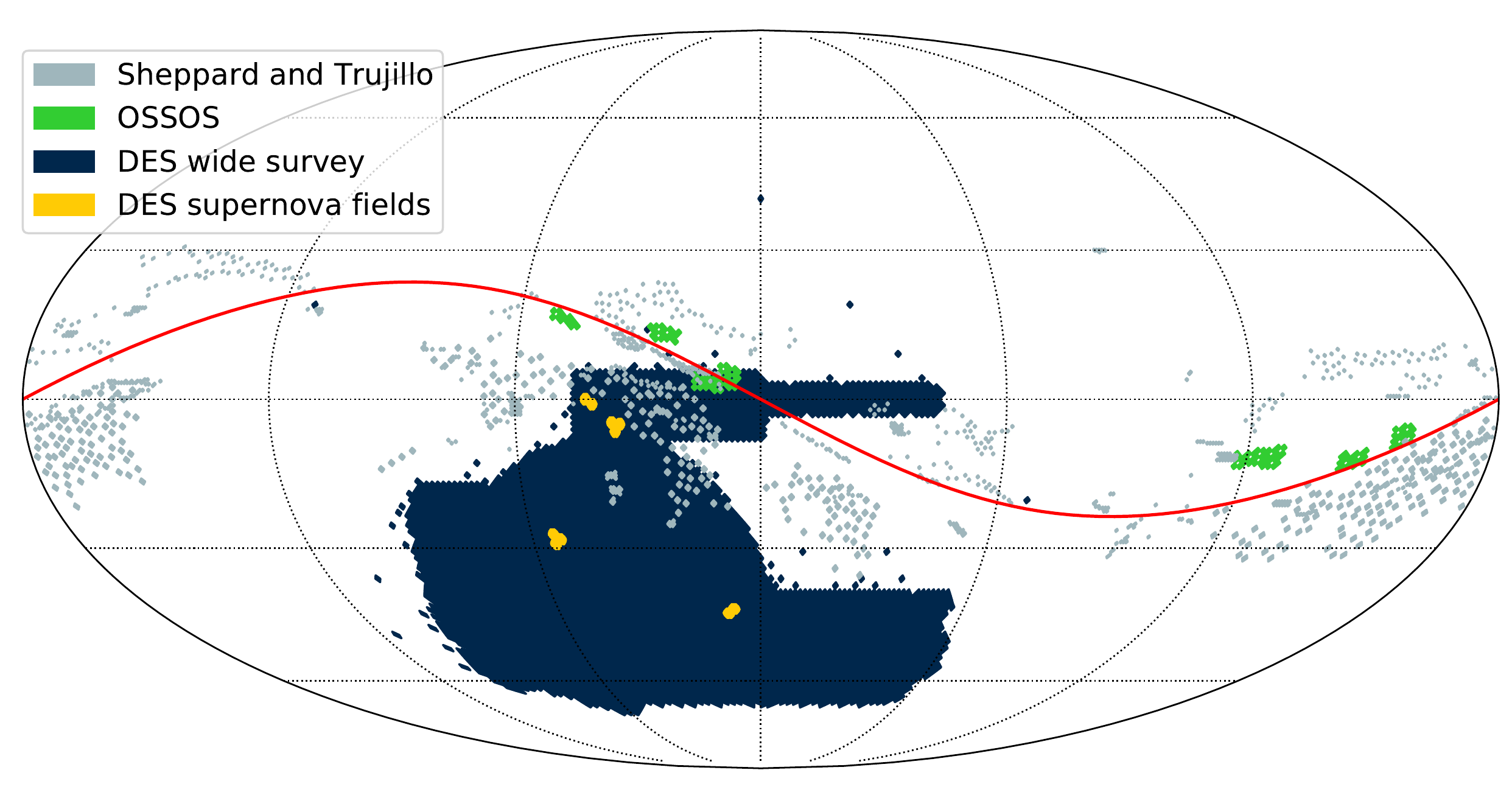}
    \caption{A HEALPix mapping of the currently-released sky coverage of the three major TNO surveys of this generation. The surveys by OSSOS and Sheppard and Trujillo hug the ecliptic plane (plotted in red), while DES, designed as a cosmological survey, has a much more expansive footprint.}
    \label{fig:pointings}
\end{figure}

The three surveys we consider have very different designs and scientific goals, and consequently quite different ETNO selection functions. This is readily apparent from their survey footprints, shown in Figure~\ref{fig:pointings}. DES, which was on-sky between 2012 and 2019, used the Dark Energy Camera (DECam, \citet{2015AJ....150..150F}) on the 4-meter Blanco telescope at CTIO to carry out an extragalactic survey designed to measure cosmological parameters. It consisted of two interwoven surveys. In the 30 sq.\ deg.\ supernova survey, ten separate fields were visited approximately weekly in the $griz$ bands during the six months per year that DES was in operation. In the 5000 sq.\ deg.\ wide survey, each field was imaged a total of 10 times at a sparse temporal cadence in each of the $grizY$ bands over the duration of the survey. The wide survey reached a limiting $r$-band magnitude of $\approx 23.5$. DES had limited near-ecliptic coverage centered near ecliptic longitude of 0, and a large off-ecliptic footprint that made it particularly sensitive to high-inclination objects. For our main analysis, we consider only the ETNOs detected in the DES wide survey, and treat the supernova fields separately. The OSSOS survey (2013-2017), by contrast, was optimized to detect and track TNOs in eight $\sim20$~sq.\ deg.\ blocks distributed along the ecliptic. This survey used the 3.6-meter Canada-France-Hawaii Telescope and reached a limiting $r$-band magnitude of 24.1–25.2. Finally, the ST survey (2007-2015) used the Blanco, Subaru, Large Binocular, and Magellan telescopes to cover 1080 sq.\ deg.\  at an average distance of 13 deg.\ from the ecliptic to a depth of approximately $VR\sim 25$. This survey aimed to detect the most distant objects: ETNOs and Inner Oort-Cloud (IOC) objects such as Sedna. Therefore, only those candidates with an estimated heliocentric distance greater than 50~au were selected for followup and tracking. 

The most complete way to account for survey bias in the discovery of the solar system objects is to use a survey simulator \citep{2011AJ....142..131P, 2018arXiv180200460L}. In essence, a survey simulator simulates detections of a model population of solar system bodies by using a survey's pointing history, depth, and tracking criteria. This allows for the computation of a survey's selection function for a given population, which enables us to account for bias, and therefore understand the true underlying populations. While it gives a reasonable approximation, the technique employed in BB19 cannot fully substitute for actually simulating each survey to calculate its selection function. Since the known ETNOs were discovered by a variety of surveys, the task of developing an appropriate simulator is nontrivial. Our simulator (\texttt{FastSSim}) is highly parametric, requiring only the few pieces of information common among all well-characterized surveys: pointing history, limiting magnitudes, and followup criteria.\footnote{The tools for the \texttt{FastSSim} algorithm have now been compiled into an open-source Python package \texttt{SpaceRocks}. It is under active development at \href{https://github.com/kjnapier/spacerocks}{https://github.com/kjnapier/spacerocks}.} The basic flow of the simulator is as follows:

\begin{enumerate}
    \item Map a survey's published pointing history to a HEALPix\footnote{It is not important that we used a HEALPix mapping. We could have used any mapping onto the sphere.} grid, as in Figure~\ref{fig:pointings}.
    \item Generate a distribution of fake objects at a single epoch.
    \item Calculate the objects' HEALPix pixels and apparent magnitudes.
    \item Determine which fake objects fall in a survey's footprint.
    \item Make cuts according to the survey's limiting magnitudes and followup criteria. 
\end{enumerate}

Note that this simulation method makes several approximations. We compute the sky coordinates of our objects at objects at a single epoch, we use a single color and limiting magnitude for each survey field, we do not consider CCD-level detections (so we do not account for complications such as chip gaps), and we employ a step-function detection criterion (so we do not model survey cadence or linking efficiency). We use a single HEALPix pixel for each survey pointing. We have chosen the pixel scales for each telescope as follows: Blanco uses NSIDE of 64 (except for the DES supernova fields, for which we use NSIDE of 1024), and the Magellan, Large Binocular, and Subaru telescopes use NSIDE of 128. These assumptions ignore the time history of the surveys, as well as apparent motion of the objects. \texttt{FastSSim} works well for this application because the objects move slowly, the telescopes have large fields of view, and the sensitivity does not have much spatial variation.

\begin{figure}[ht]
    \centering
    \includegraphics[width=0.8\textwidth]{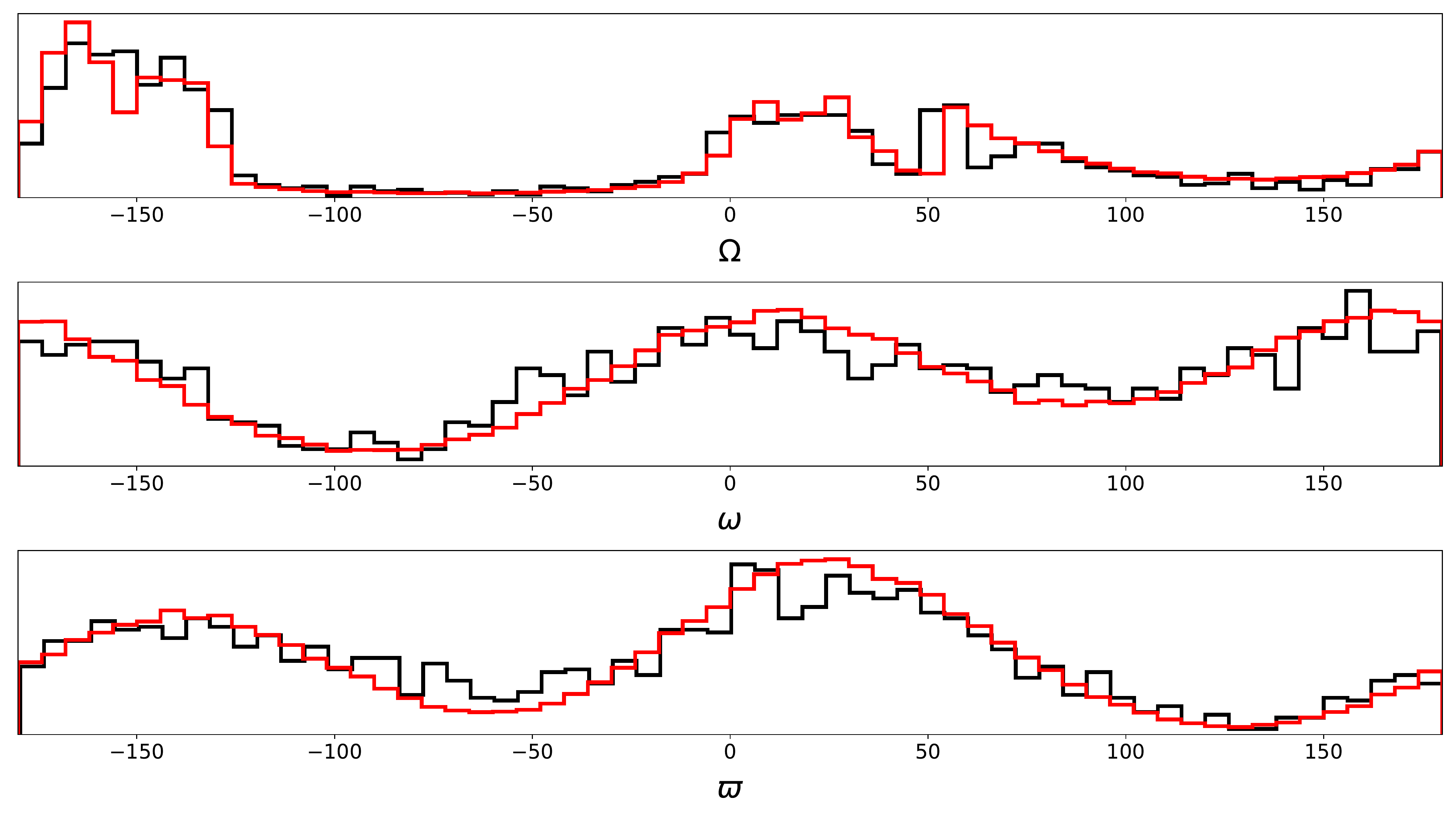}
    \caption{OSSOS selection functions for 1,615 detections from the nominal population described in \S 3 in the angles $\omega$, $\Omega$, and $\varpi$, as calculated by \texttt{FastSSim} (red) and the OSSOS/CFEPS survey simulator (black). The distributions have the same general shape.}
    \label{fig:comparison} 
\end{figure}

We acquired the non-DES survey pointings and limiting magnitudes from \citet{ST16}, \citet{Sheppard_2019}, and \citet{2018ApJS..236...18B}. We choose each HEALPix pixel size to  most closely match the field of view of the telescope used. This does not allow for a perfect mapping between pointings and pixels, but it turns out to be sufficient for our needs. In fact, we find that \texttt{FastSSim} performs remarkably well in cross-checks against both our full chip-level DES simulator \citep{HamiltonPhD}, and the OSSOS survey simulator described in \citet{2011AJ....142..131P} (see Figure \ref{fig:comparison}). While \texttt{FastSSim} misses some of the fine details of the selection functions, the small sample of ETNOs and the approximate nature of this analysis make such fine details unimportant to our overall conclusion. Given the success of these cross-checks, we are confident in extending its use to characterize the survey of Sheppard and Trujillo.

To simulate the surveys we randomly generate ETNOs in accordance with a nominal scattered disk model (specified in \S\ref{sec:SS_model}) until each survey has accumulated $10^5$ detections, to allow for high-resolution characterization of a survey's sensitivity in $\varpi$ and $\Omega$. This typically requires the generation of approximately $10^{10}$ fakes, so our set of simulated objects spans the parameter space of the ETNOs. We consider an object to be detected if it is in one of the survey's HEALPix pixels, is brighter than the pixel's limiting magnitude, and has a perihelion distance $q \geq$ 30~au. For the survey of Sheppard and Trujillo, we satisfy a tracking criterion specified in \citet{2016ApJ...825L..13S} by requiring an object to have a heliocentric distance of at least 50~au at the time of detection.

As a quantitative example of the effectiveness of \texttt{FastSSim}, Figure \ref{fig:comparison} shows a comparison with the CFEPS/OSSOS simulator. For this test each simulator uses the population model defined in \S 3. Using Kuiper's test, we find that the distributions of the 1,615 data points calculated by \texttt{FastSSim} are statistically indistinguishable from those computed using the CFEPS/OSSOS survey simulator. Thus in order to distinguish the two simulators one would need $>1,615$ ETNOs---well above the quantity discovered by OSSOS. We achieve similar results in quantitative comparisons against the distributions computed using the DES survey simulators (see Figure 5.1 in \cite{HamiltonPhD} and Figure 2 in \cite{Bernardinelli2020}).

\begin{figure}[ht]
    \centering
    \includegraphics[width=0.7\textwidth]{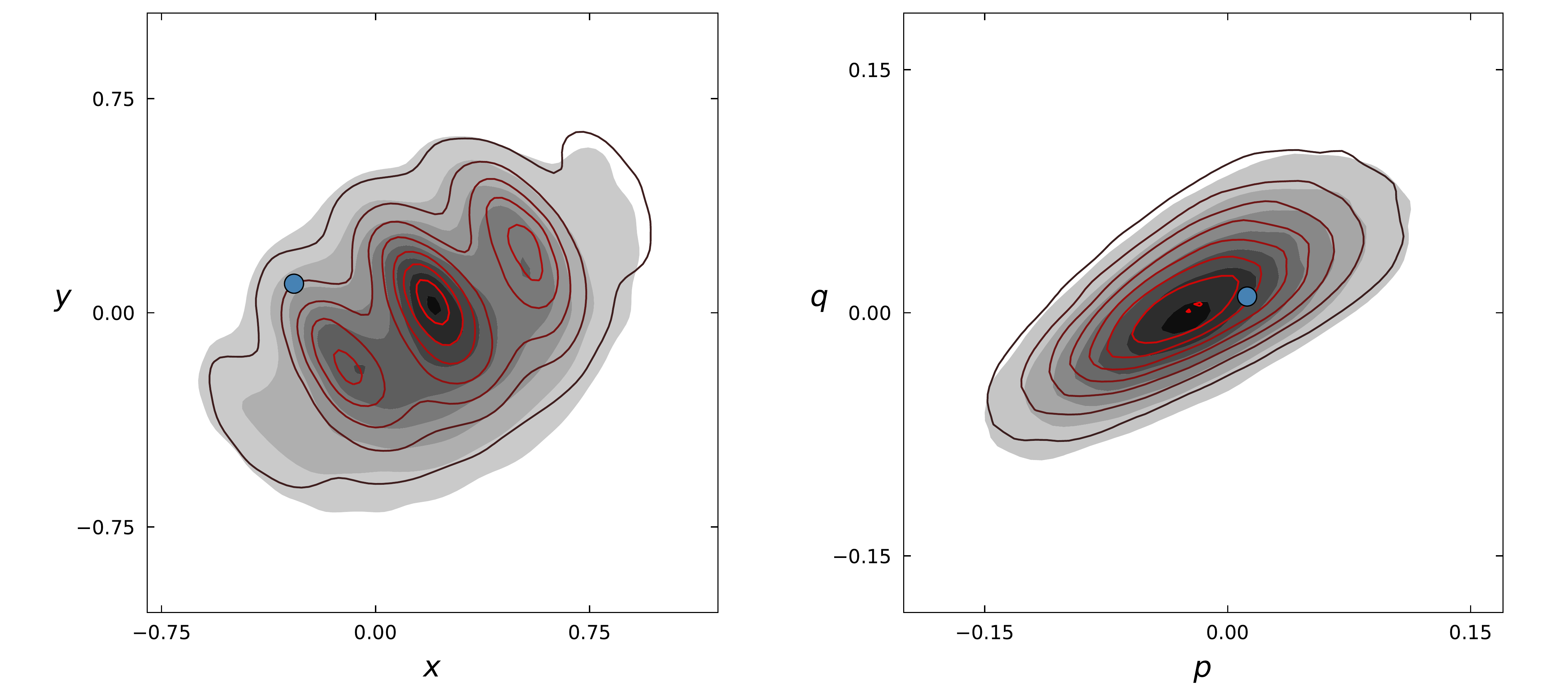}
    \caption{The black contours (which follow a linear scale) are Gaussian kernel density estimates of $10^6$ iterations of sampling 5 points from the OSSOS biases calculated using the CFEPS/OSSOS survey simulator (because OSSOS discovered 5 ETNOs), and plotting the mean $(x,y)$ and $(p, q)$ positions (see \S 4 for definitions of $x$, $y$, $p$, and $q$). The red contours represent the same statistic, calculated using {\texttt{FastSSim}}. The blue points are the mean positions of the 5 ETNOs discovered by OSSOS. Our simulator reproduces the results of the OSSOS simulator with better fidelity than the heuristic method of BB19 (see their Figure 4).}
    \label{fig:OSSOS_xy_comparison} 
\end{figure}

In Figure \ref{fig:OSSOS_xy_comparison} we plot a Gaussian kernel density estimate of each simulator's detections in the $(x, y)$ and $(p, q)$ spaces ($x$, $y$, $p$, and $q$ are defined in \S 4). Since the distributions appear to be in good agreement we believe that our selection functions, which are derived directly from the surveys' pointing histories, depths, and TNO tracking strategies, more faithfully model their respective surveys than than those inferred indirectly in BB19 (see their Figure 4). Furthermore, the p-values we calculate using the CFEPS/OSSOS survey simulator do not significantly differ from those we calculate using {\texttt{FastSSim}}.

\section{Scattered Disk Model}
\label{sec:SS_model}
To test the dependence of our analysis on the choice of scattered disk model, we simulated models with various distributions of semi-major axis ($a$), eccentricity ($e$), inclination ($i$), and absolute magnitude ($H$), while keeping the orbital angles $\Omega$ and $\omega$ (and thus $\varpi$) uniform from $0\degree$ to $360\degree$. We tested manifold permutations with the parameter distributions: $N(a) \propto a^\zeta$ with $a \in [230\,\text{au}, 1600\,\text{au}]$ and $\zeta \in [0.5, 1.0]$,  uniform $i \in [0\degree, 60\degree]$, Brown distribution $i$ (\citet{Brown2001}) with a variety of widths ranging from 5$\degree$ to 25$\degree$, and $N(H) \propto 10^{H\zeta}$ with $H \in [4, 10]$ and $\zeta \in [0.6, 0.9]$.

We found that our conclusions were not significantly affected by the variation of the model parameters. Our results are also robust to changes in pericenter distribution (see the Appendix for the distribution of orbital elements for populations with $q > 30, 35$, and $38$ au). \citep{OSSOS17} found similar resilience to changes in scattered disk model. Noting the weak dependence of the outcome of our simulations on the choice of model, we proceed using the following scattered disk model:

\begin{itemize}
    \item[--] $a$ follows a single power-law distribution such that $N(a) \propto a^{0.7}$, where $a \in [230\,\text{au}, 1600\,\text{au}]$
    \item[--] $e$ is distributed uniformly $\in [0.69, 0.999]$
    \item[--] $i$ follows a Brown distribution such that $N(i) \propto \sin(i) \exp{\left[-\frac{(i-\mu_i)^2}{2\sigma_i^2}\right]}$ with $\mu_i = 0 \degree$ and $\sigma_i = 15 \degree$
    \item[--] $H$ follows a single power-law distribution such that $N(H) \propto 10^{0.8 H}$, where $H \in [4, 10]$
    \item[--] Perihelion distance $q > 30\,\text{au}$ 
\end{itemize}

These model parameters produce posteriors in $a$, $e$, $q$, $i$, and $H$ that appear to be in reasonable agreement with the real ETNO detections by each survey. See the Appendix for histograms of the posteriors in each of these variables, overlaid with a rug plot of each survey's real detections.

\section{Analysis and Results}
\label{sec:analysis}

\begin{figure}[ht]
    \centering
    \includegraphics[width=1\textwidth]{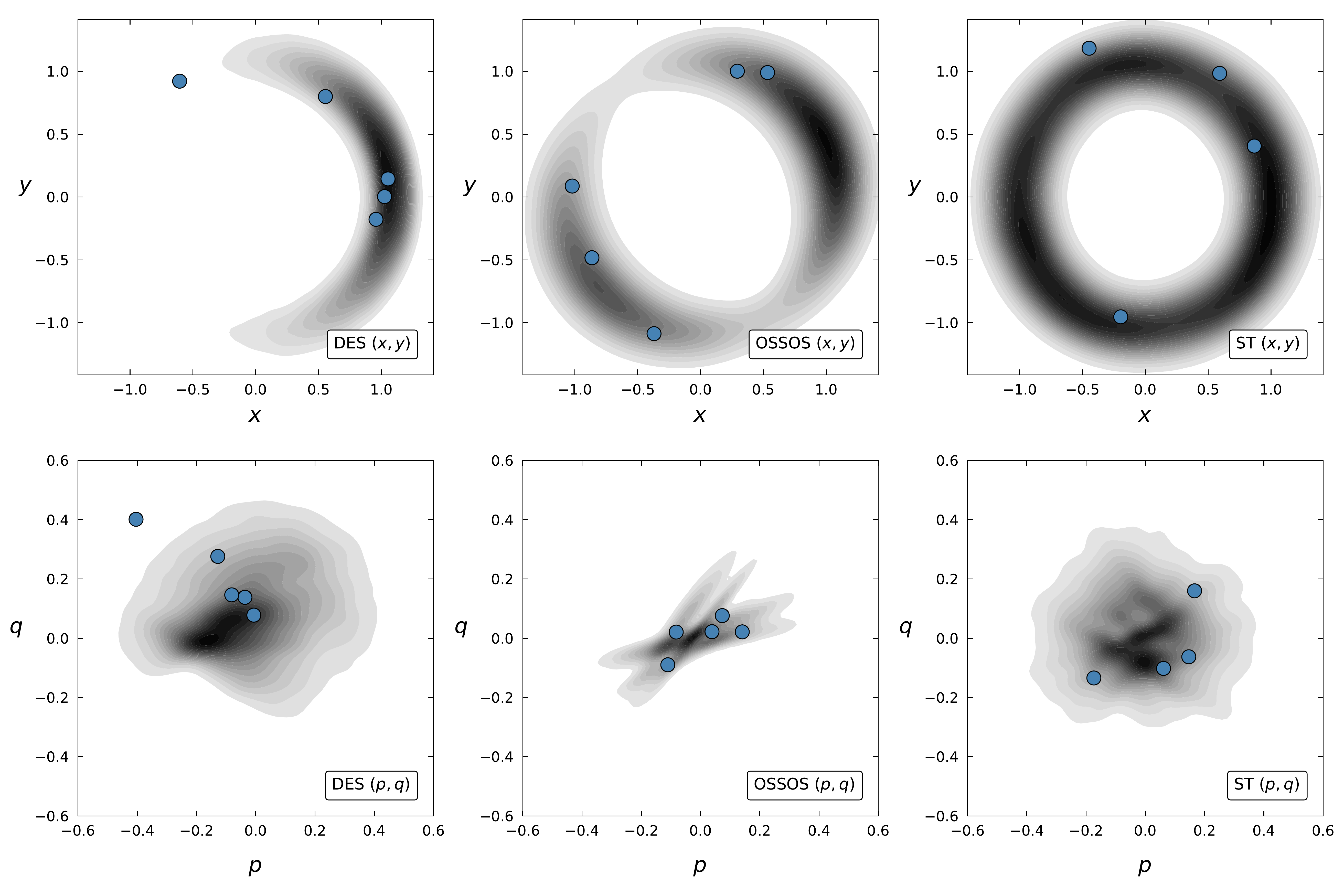}
    \caption{Kernel density estimates of each survey's selection function in the canonical $xy$-space (top row) and $pq$-space (bottom row). The contours represent simulated detections (the contours scale linearly, and darker contours are more densely populated), while the blue dots represent the ETNOs detected by each survey. The outlier in both DES panels is the object 2015 BP$_{519}$ \citep{Becker_2018}.}
    \label{fig:canonical}
\end{figure}

Performing a clustering analysis in the variables $\varpi$ and $\Omega$ is complicated, as the two are strongly correlated. We proceed by working in the orthogonal $\{x,y,p,q\}$ basis discussed in BB19 (importantly, $\varpi$ and $\Omega$ are linearly independent in this basis). Note that these vectors are not normalized, but instead have their lengths modulated by eccentricity and inclination. The coordinates are defined as follows. 
\begin{equation}
    \Gamma = 1-\sqrt{1-e^2} \qquad \qquad Z = \sqrt{1-e^2}\left[ 1 - \cos (i) \right]
\end{equation}
\begin{equation}
    x = \sqrt{2 \Gamma} \cos(\varpi) \qquad\qquad  y = \sqrt{2 \Gamma} \sin(\varpi)
\end{equation}
\begin{equation}
    p = \sqrt{2 Z} \cos(\Omega) \qquad\qquad  q = \sqrt{2 Z} \sin(\Omega)
\end{equation}
Note that $\Gamma$ and $Z$ have been scaled by a factor of $\sqrt{G M_{\odot}a}$ from their traditional forms, since the semi-major axis is not relevant to this argument. Figure \ref{fig:canonical} shows our calculated selection functions in the $xy$ and $pq$-planes.

\begin{figure}[ht]
    \centering
    \includegraphics[width=1\textwidth]{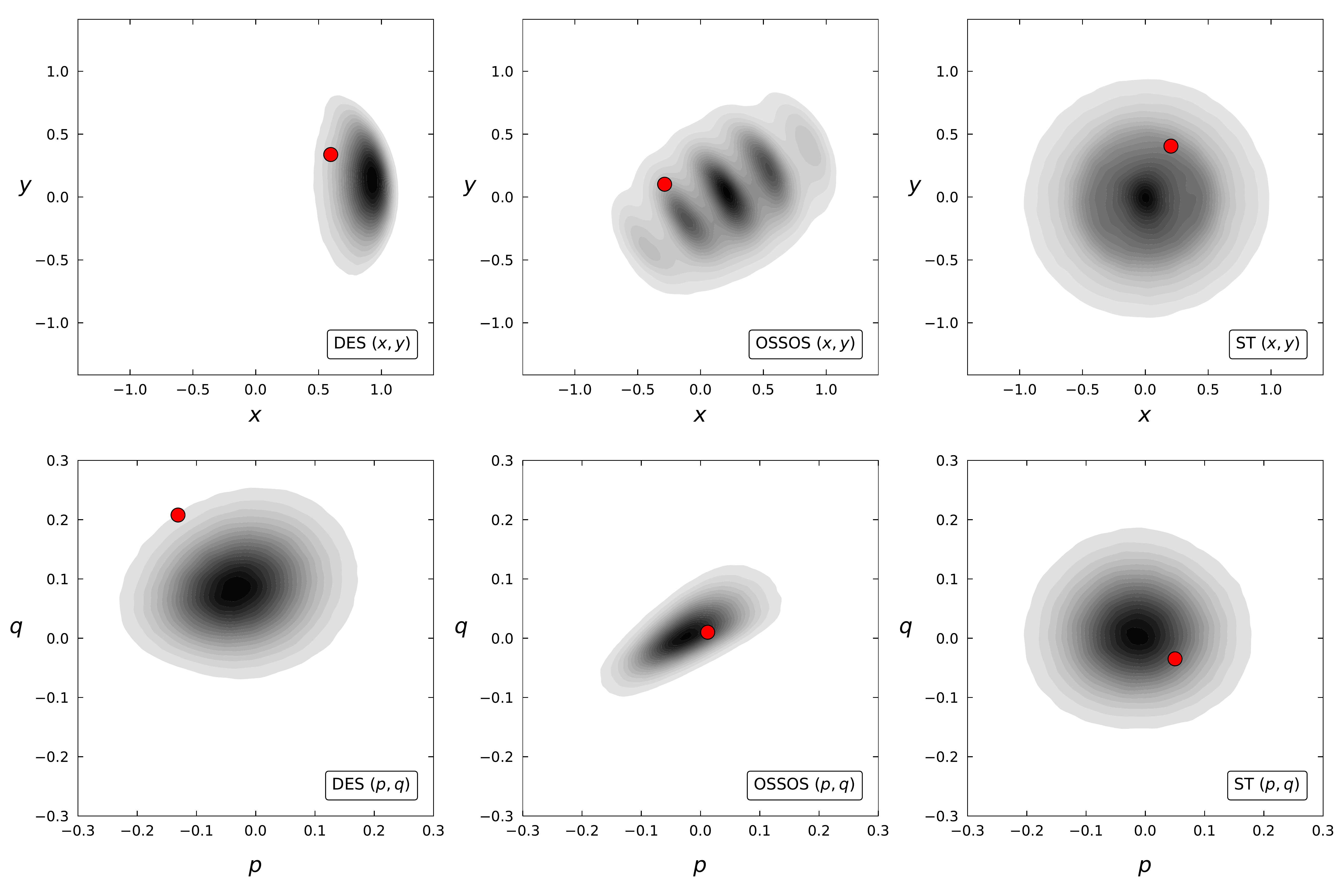}
    \caption{Kernel density estimates of the mean $(x, y)$ and $(p, q)$ position of $10^6$ samples of ETNOs drawn from the PDFs shown in Figure \ref{fig:canonical}. The number of objects in each sample corresponds to the number of ETNOs detected by the given survey. The contours represent the samples (the contours scale linearly, and darker contours are more densely populated), while the red dots represent the mean position of the ETNOs detected by each survey.}
    \label{fig:mean_xypq}
\end{figure}

For the sake of comparison, we used the method presented in BB19 to test the consistency of each survey's detected ETNOs with its selection function. We first perform $10^6$ iterations sampling from our simulated detections a set of objects whose cardinality is equal to that of the set of real ETNOs detected by the given survey. We then take the average $\{x,y,p,q\}$ position of each sample, and use these values to construct a 4-dimensional histogram. We display a Gaussian kernel density estimation of these data in the $xy$ and $pq$-planes in Figure \ref{fig:mean_xypq}. 

We perform a Gaussian kernel density estimation on our mean-sampled histograms to obtain a probability distribution function (PDF). Next we draw N samples from our simulated data (where N is the number of ETNOs actually detected by the survey), find the mean $\{x, y, p, q\}$ position, and evaluate our PDF at that position. We repeat this $10^5$ times to construct a likelihood function. Next we compute this value for the ETNOs actually discovered by the survey. To calculate the probability of a survey detecting the ETNOs it actually detected (as opposed to some other set of ETNOs), we find the fraction of the $10^5$ sample likelihood values that the survey's actual likelihood value exceeds. Rounded to the nearest $1\%$, this probability for each survey is as follows: ${\mathcal{P}_{DES}} \sim 0.06$, ${\mathcal{P}_{OSSOS}} \sim 0.53$, and ${\mathcal{P}_{ST}} \sim 0.59$.

The joint probability of N surveys detecting objects with given probabilities (or some less likely set of values) can be calculated as the volume under the surface of constant product of probabilities in the domain of the N-dimensional unit hypercube, given by 
\begin{equation}
    \mathcal{P}_{joint} = P \sum_{k=0}^N (-1)^k \frac{\log(P)^k}{k!}
    \label{eq:joint_prob}
\end{equation} 
\noindent where $P \equiv \prod_{k} \mathcal{P}_{k}$. In our case, $k \in \{\text{DES}, \text{OSSOS}, \text{ST}\}$. Using Equation \ref{eq:joint_prob}, we calculate the joint probability to be $24\%$.

With such a small sample size, this work is sensitive to outliers and the definition of the ETNOs itself. The high-inclination object 2015 BP$_{519}$ is among the most dynamically anomalous objects in the solar system \citep{Becker_2018}, and we cannot discount the possibility that it is of a different dynamical origin than the other ETNOs. If we redo our analysis without 2015 BP$_{519}$, $\mathcal{P}_{DES}$ increases to $84\%$, and thus $\mathcal{P}_{joint}$ increases to $\sim 85\%$. 2014 FE$_{72}$ has an extremely large semi-major axis---roughly four standard deviations above the mean of the ETNOs considered in this work. Its large semi-major axis carries it deep into the IOC region, where interactions with galactic tides make its secular relationship with a putative Planet X/Planet 9 less certain. If we exclude 2014 FE$_{72}$, $\mathcal{P}_{\text{ST}}$ increases to $88\%$ and thus $\mathcal{P}_{\text{joint}}$ increases to $31\%$. If we include 2012 VP$_{113}$, $\mathcal{P}_{\text{ST}}$ increases to $60\%$, and $\mathcal{P}_{\text{joint}}$ remains $24\%$. We also address the fact that the clustering by a putative Planet~X/Planet~9 should be more robust in the sample of ETNOs with $q > 40$ au, since these objects avoid strong perturbations by Neptune. If we restrict our ETNOs to these 8 objects, $\mathcal{P}_{\text{joint}}$ increases to $94\%$. Finally, we analyze the subset of objects which are either stable or metastable in the presence of the putative Planet~X/Planet~9 \citep{P9Review}: 2015 TG$_{387}$, 2013 SY$_{99}$, 2015 RX$_{245}$, 2014 SR$_{349}$, 2012 VP$_{113}$, 2013 RA$_{109}$, and 2013 FT$_{28}$. For this subset $\mathcal{P}_{\text{joint}} = 82\%$.

For the sake of completeness, we also use a more traditional sampling method to determine the significance of the clustering of ETNOs. We begin by performing a Gaussian kernel density estimate on each survey's posterior distributions. We then perform $10^5$ iterations in which we randomly draw $N$ points from each survey's posterior distribution (where $N$ is the number of ETNOs detected by the survey) and multiply each of the $N$ probabilities together to calculate a likelihood. Finally, we calculate the same metric for each survey's actual detections and compare the value to the distribution of our samples. As before, the probability for each survey is the fraction of the $10^5$ sample likelihood values that the survey's actual likelihood value exceeds. Rounded to the nearest $1\%$, the probability for each survey is as follows: ${\mathcal{P}_{DES}} \sim 0.06$, ${\mathcal{P}_{OSSOS}} \sim 0.41$, and ${\mathcal{P}_{ST}} \sim 0.43$. The joint probability is thus $17\%$.

\begin{figure}[h]
    \centering
    \includegraphics[width=0.7\textwidth]{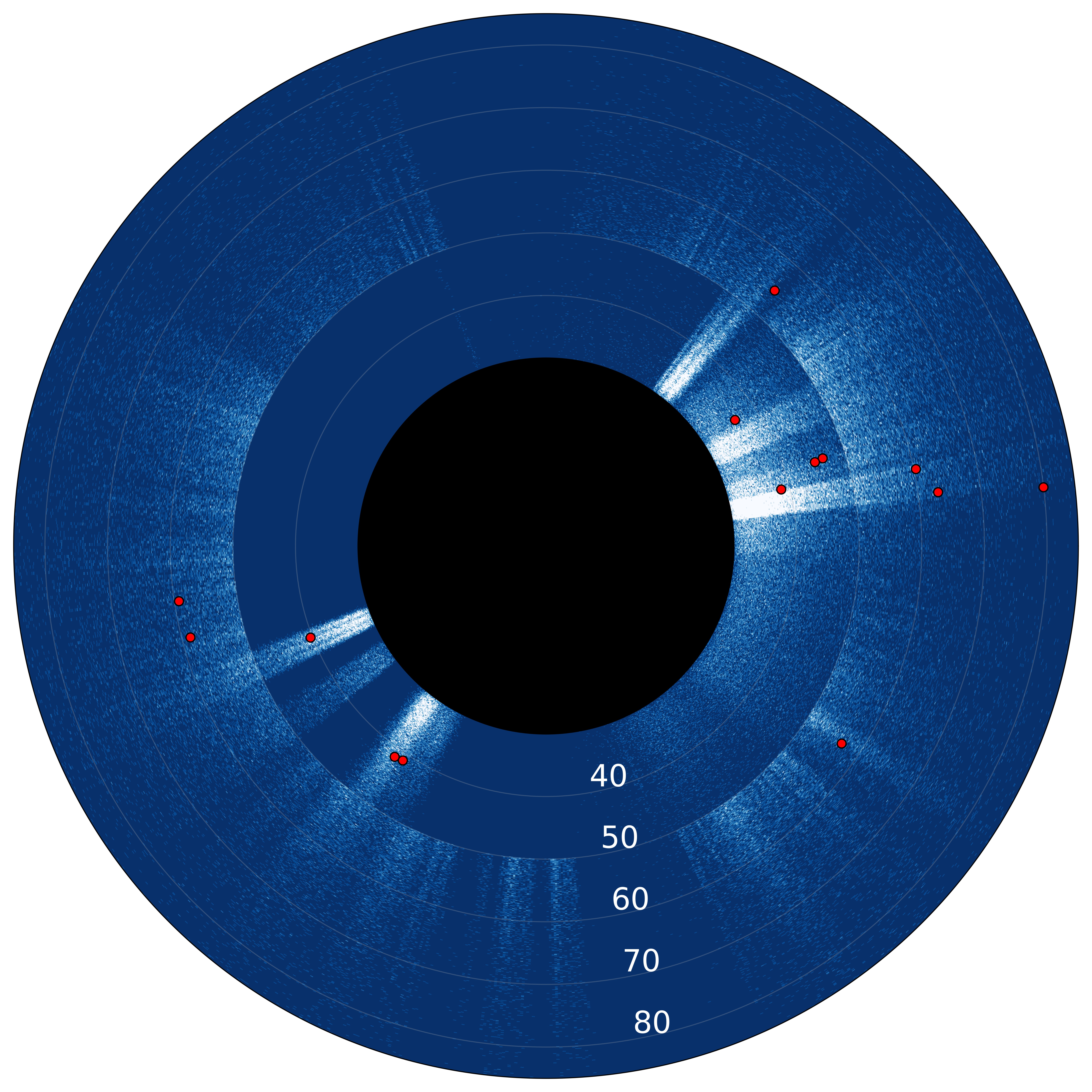} 
    \caption{Combined ETNO selection function for all three surveys. The radial quantity is the ETNO's barycentric distance, and the azimuthal quantity is true longitude. The edge of the black circle is at 30 au. The white regions represent the combined surveys' sensitivity (brighter regions correspond to higher sensitivity), weighted by the number of real ETNO detections. The red dots represent the real ETNOs at the epoch of discovery. The outer ring is caused by the 50 au tracking criterion imposed by ST.}
    \label{fig:corona}
\end{figure}

For a more physically intuitive representation of the survey bias, refer to Figure \ref{fig:corona}. Here the radial quantity represents the barycentric distance, and the azimuthal quantity represents true longitude (the true anomaly + $\varpi$). The edge of the black circle is at 30 au. The white regions represent the combined surveys' sensitivity (brighter regions correspond to higher sensitivity), weighted by the number of real ETNO detections. The red dots represent the real ETNOs at the epoch of discovery. The observations are in good agreement with the combined selection function, qualitatively confirming the conclusions of our formal statistical analysis performed on canonical variables.  

\subsection{DES Supernova Fields}

\begin{figure}[ht]
    \centering
    \includegraphics[width=0.8\textwidth]{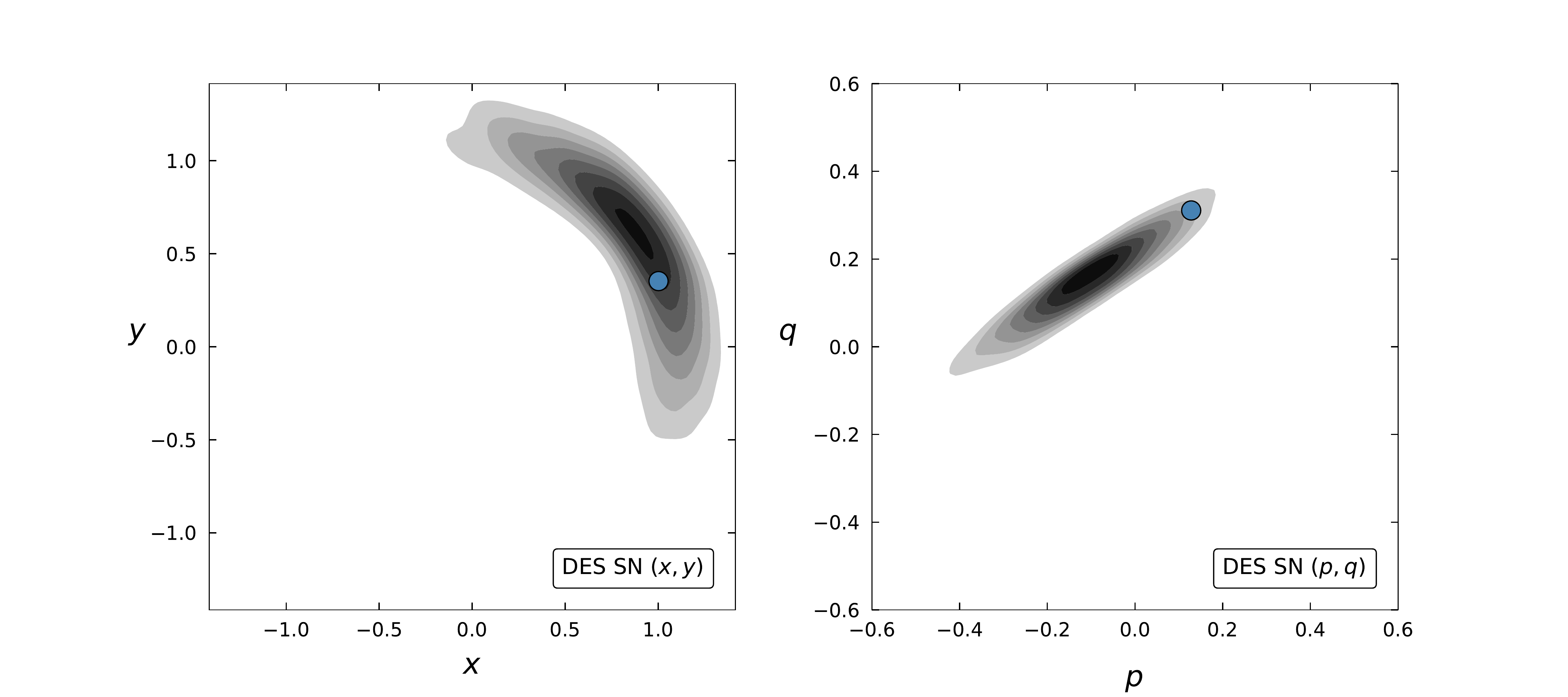} 
    \caption{Kernel density estimates of the DES SN selection function in the canonical $xy$-space (left) and $pq$-space (right). The contours represent simulated detections (the contours scale linearly, and darker contours are more densely populated), while the blue dots represent 2013 RF$_{98}$.}
    \label{fig:rf_xypq}
\end{figure}

The ETNO 2013 RF$_{98}$ was discovered in the deep DES supernova fields (DES SN hereafter). Since the DES SN fields are so small, they suffer from severe selection bias. Additionally, since their observing cadence and depth ($\sim$24.5 in the $r$-band) are significantly different than the wide survey, they need to be treated independently. We generated 1,829 simulated detections in the DES SN fields (since the fields are so small, it is computationally prohibitive to generate $10^5$ synthetic detections as we do for DES, OSSOS, and ST) from the population model defined in \S 3. We show the posteriors in $\{a, e, i, H, \Omega, \varpi\}$ in Figures \ref{fig:q_post}, \ref{fig:a_post}, \ref{fig:e_post}, \ref{fig:inc_post}, \ref{fig:H_post}, \ref{fig:node_post}, and \ref{fig:varpi_post}. Figure \ref{fig:rf_xypq} shows a kernel density estimate of the detections in $\{x, y, p, q\}$ space. In all parameters, 2013 RF$_{98}$ appears to be a rather ordinary detection for the DES SN fields.  

Since there is only one data point here, we can just numerically integrate to find $\mathcal{P}_{\text{DES SN}} = 0.33$ (i.e. a p-value of 0.33). Treating DES SN as its own survey, we may use Equation \ref{eq:joint_prob} to calculate the 4-survey joint probability to find $\mathcal{P}_{\text{joint}} = 25\%$.

\section{Discussion and Conclusions}
\label{sec:discussion}

We use quantified selection bias calculations on all the ETNOs discovered by the three most productive ETNO surveys, each with with a quite different survey strategy and selection function, to test the consistency of the ETNOs with a uniform underlying distribution. Given a joint probability between $17\%$ and $94\%$ (i.e. a $p$-value between 0.17 and 0.94), we conclude that the sample of ETNOs from well-characterized surveys is fully consistent with an underlying parent population with uniform distributions in the longitudes $\varpi$ and $\Omega$. Our result differs drastically from the corresponding value in BB19 of $0.2\%$. Closer inspection sheds some light on the apparent discrepancy. If we examine only the overlapping set of ETNOs used this work and in BB19 (2015 BP$_{519}$, 2013 RF$_{98}$, 2013 SY$_{99}$, 2015 RX$_{245}$, 2015 GT$_{50}$, 2015 KG$_{163}$, 2013 FT$_{28}$, 2014 SR$_{349}$, and 2014 FE$_{72}$), $\mathcal{P}_{\text{joint}}$ drops to $< 0.005$. This indicates an expected issue: small number statistics are sensitive to fluctuations. For example, when BB19 performed their analysis a small but important set of the ETNOs had not yet been reported to the MPC. As a concrete demonstration of the importance of the omission of a few ETNOs from BB19, consider DES. Of the five ETNOs discovered by the DES wide survey, BB19 included only 2015 BP$_{519}$. From Figure~\ref{fig:canonical} it is clear that this object lands in an extremely low-probability region. This drives down $\mathcal{P}_{\text{joint}}$, and thus gives a satisfactory answer as to why the result of this work differs so significantly from that of BB19.

It is important to note that our work does not explicitly rule out Planet~X/Planet~9; its dynamical effects are not yet well enough defined to falsify its existence with current data. This work also does not analyze whether some form of clustering could be consistent with the 14 ETNOs we consider. For example, the ETNOs could happen to be clustered precisely where current surveys have looked. In that case, a survey with coverage orthogonal to the regions shown in Figure \ref{fig:corona} would find far fewer ETNOs than expected. Various realizations of Planet~X/Planet~9 predict clustering of various widths, modalities, and libration amplitudes and frequencies; we do not test for consistency with any of these distributions. Instead, we have shown that given the current set of ETNOs from well-characterized surveys, there is no evidence to rule out the null hypothesis. Increasing the sample of ETNOs with ongoing and future surveys with different selection functions such as the Deep Ecliptic Exploration Project (DEEP) \citep{DEEP2019} and the Legacy Survey of Space and Time (LSST) at the Vera Rubin Observatory \citep{LSSTsolarsystem} will allow for more restrictive results. Despite other lines of indirect evidence for Planet~X/Planet~9, in the absence of clear evidence for clustering of the ETNOs the argument becomes much weaker. Future studies should consider other mechanisms capable of giving the outer solar system its observed structure, while preserving a uniform distribution of ETNOs in the longitudes $\Omega$ and $\varpi$.

\acknowledgements 
This material is based upon work supported by the National Aeronautics and Space
Administration under Grant No. NNX17AF21G issued through the SSO Planetary Astronomy Program and by the National Science Foundation under Grant No. AST-2009096.

Funding for the DES Projects has been provided by the U.S. Department of Energy, the U.S. National Science Foundation, the Ministry of Science and Education of Spain, 
the Science and Technology Facilities Council of the United Kingdom, the Higher Education Funding Council for England, the National Center for Supercomputing 
Applications at the University of Illinois at Urbana-Champaign, the Kavli Institute of Cosmological Physics at the University of Chicago, 
the Center for Cosmology and Astro-Particle Physics at the Ohio State University,
the Mitchell Institute for Fundamental Physics and Astronomy at Texas A\&M University, Financiadora de Estudos e Projetos, 
Funda{\c c}{\~a}o Carlos Chagas Filho de Amparo {\`a} Pesquisa do Estado do Rio de Janeiro, Conselho Nacional de Desenvolvimento Cient{\'i}fico e Tecnol{\'o}gico and 
the Minist{\'e}rio da Ci{\^e}ncia, Tecnologia e Inova{\c c}{\~a}o, the Deutsche Forschungsgemeinschaft and the Collaborating Institutions in the Dark Energy Survey. 

The Collaborating Institutions are Argonne National Laboratory, the University of California at Santa Cruz, the University of Cambridge, Centro de Investigaciones Energ{\'e}ticas, 
Medioambientales y Tecnol{\'o}gicas-Madrid, the University of Chicago, University College London, the DES-Brazil Consortium, the University of Edinburgh, 
the Eidgen{\"o}ssische Technische Hochschule (ETH) Z{\"u}rich, 
Fermi National Accelerator Laboratory, the University of Illinois at Urbana-Champaign, the Institut de Ci{\`e}ncies de l'Espai (IEEC/CSIC), 
the Institut de F{\'i}sica d'Altes Energies, Lawrence Berkeley National Laboratory, the Ludwig-Maximilians Universit{\"a}t M{\"u}nchen and the associated Excellence Cluster Universe, 
the University of Michigan, NFS's NOIRLab, the University of Nottingham, The Ohio State University, the University of Pennsylvania, the University of Portsmouth, 
SLAC National Accelerator Laboratory, Stanford University, the University of Sussex, Texas A\&M University, and the OzDES Membership Consortium.

Based in part on observations at Cerro Tololo Inter-American Observatory at NSF’s NOIRLab (NOIRLab Prop. ID 2012B-0001; PI: J. Frieman), which is managed by the Association of Universities for Research in Astronomy (AURA) under a cooperative agreement with the National Science Foundation.

The DES data management system is supported by the National Science Foundation under Grant Numbers AST-1138766 and AST-1536171.
The DES participants from Spanish institutions are partially supported by MICINN under grants ESP2017-89838, PGC2018-094773, PGC2018-102021, SEV-2016-0588, SEV-2016-0597, and MDM-2015-0509, some of which include ERDF funds from the European Union. IFAE is partially funded by the CERCA program of the Generalitat de Catalunya.
Research leading to these results has received funding from the European Research
Council under the European Union's Seventh Framework Program (FP7/2007-2013) including ERC grant agreements 240672, 291329, and 306478.
We  acknowledge support from the Brazilian Instituto Nacional de Ci\^encia
e Tecnologia (INCT) do e-Universo (CNPq grant 465376/2014-2).

This manuscript has been authored by Fermi Research Alliance, LLC under Contract No. DE-AC02-07CH11359 with the U.S. Department of Energy, Office of Science, Office of High Energy Physics.

\bibliographystyle{aasjournal}
\bibliography{refs}

\begin{thebibliography}{}
\expandafter\ifx\csname natexlab\endcsname\relax\def\natexlab#1{#1}\fi
\providecommand{\url}[1]{\href{#1}{#1}}
\providecommand{\dodoi}[1]{doi:~\href{http://doi.org/#1}{\nolinkurl{#1}}}
\providecommand{\doeprint}[1]{\href{http://ascl.net/#1}{\nolinkurl{http://ascl.net/#1}}}
\providecommand{\doarXiv}[1]{\href{https://arxiv.org/abs/#1}{\nolinkurl{https://arxiv.org/abs/#1}}}

\bibitem[{{Bannister} {et~al.}(2016){Bannister}, {Kavelaars}, {Petit},
  {Gladman}, {Gwyn}, {Chen}, {Volk}, {Alexandersen}, {Benecchi}, {Delsanti},
  {Fraser}, {Granvik}, {Grundy}, {Guilbert-Lepoutre}, {Hestroffer}, {Ip},
  {Jakubik}, {Jones}, {Kaib}, {Kavelaars}, {Lacerda}, {Lawler}, {Lehner},
  {Lin}, {Lister}, {Lykawka}, {Monty}, {Marsset}, {Murray-Clay}, {Noll},
  {Parker}, {Pike}, {Rousselot}, {Rusk}, {Schwamb}, {Shankman}, {Sicardy},
  {Vernazza}, \& {Wang}}]{Bannister2016}
{Bannister}, M.~T., {Kavelaars}, J.~J., {Petit}, J.-M., {et~al.} 2016, \aj,
  152, 70, \dodoi{10.3847/0004-6256/152/3/70}

\bibitem[{{Bannister} {et~al.}(2018){Bannister}, {Gladman}, {Kavelaars},
  {Petit}, {Volk}, {Chen}, {Alexand ersen}, {Gwyn}, {Schwamb}, {Ashton},
  {Benecchi}, {Cabral}, {Dawson}, {Delsanti}, {Fraser}, {Granvik},
  {Greenstreet}, {Guilbert-Lepoutre}, {Ip}, {Jakubik}, {Jones}, {Kaib},
  {Lacerda}, {Van Laerhoven}, {Lawler}, {Lehner}, {Lin}, {Lykawka}, {Marsset},
  {Murray-Clay}, {Pike}, {Rousselot}, {Shankman}, {Thirouin}, {Vernazza}, \&
  {Wang}}]{2018ApJS..236...18B}
{Bannister}, M.~T., {Gladman}, B.~J., {Kavelaars}, J.~J., {et~al.} 2018, \apjs,
  236, 18, \dodoi{10.3847/1538-4365/aab77a}

\bibitem[{Batygin {et~al.}(2019)Batygin, Adams, Brown, \& Becker}]{P9Review}
Batygin, K., Adams, F.~C., Brown, M.~E., \& Becker, J.~C. 2019, Physics
  Reports, 805, 1 , \dodoi{https://doi.org/10.1016/j.physrep.2019.01.009}

\bibitem[{{Batygin} \& {Brown}(2016)}]{BB16}
{Batygin}, K., \& {Brown}, M.~E. 2016, \aj, 151, 22,
  \dodoi{10.3847/0004-6256/151/2/22}

\bibitem[{Becker {et~al.}(2018)Becker, Khain, Hamilton, Adams, Gerdes,
  {et~al.}}]{Becker_2018}
Becker, J.~C., Khain, T., Hamilton, S.~J., {et~al.} 2018, The Astronomical
  Journal, 156, 81, \dodoi{10.3847/1538-3881/aad042}

\bibitem[{{Bernardinelli} {et~al.}(2020{\natexlab{a}}){Bernardinelli},
  {Bernstein}, {Sako}, {Hamilton}, {Gerdes}, {Adams}, {Saunders}, {Aguena},
  {Allam}, {Avila}, {Brooks}, {Diehl}, {Doel}, {Everett},
  {Garc{\'\i}a-Bellido}, {Gaztanaga}, {Gruendl}, {Honscheid}, {Ogando},
  {Palmese}, {Tucker}, {Walker}, {Wester}, \& {DES
  Collaboration}}]{Bernardinelli2020}
{Bernardinelli}, P.~H., {Bernstein}, G.~M., {Sako}, M., {et~al.}
  2020{\natexlab{a}}, The Planetary Science Journal, 1, 28,
  \dodoi{10.3847/PSJ/ab9d80}

\bibitem[{{Bernardinelli} {et~al.}(2020{\natexlab{b}}){Bernardinelli},
  {Bernstein}, {Sako}, {Liu}, {Saunders}, {Khain}, {Lin}, {Gerdes}, {Brout},
  {Adams}, {Belyakov}, {Somasundaram}, {Sharma}, {Locke}, {Franson}, {Becker},
  {Napier}, {Markwardt}, {Annis}, {Abbott}, {Avila}, {Brooks}, {Burke},
  {Carnero Rosell}, {Carrasco Kind}, {Castander}, {da Costa}, {De Vicente},
  {Desai}, {Diehl}, {Doel}, {Everett}, {Flaugher}, {Garc{\'\i}a-Bellido},
  {Gruen}, {Gruendl}, {Gschwend}, {Gutierrez}, {Hollowood}, {James}, {Johnson},
  {Johnson}, {Krause}, {Kuropatkin}, {Maia}, {March}, {Miquel},
  {Paz-Chinch{\'o}n}, {Plazas}, {Romer}, {Rykoff}, {S{\'a}nchez}, {Sanchez},
  {Scarpine}, {Serrano}, {Sevilla-Noarbe}, {Smith}, {Sobreira}, {Suchyta},
  {Swanson}, {Tarle}, {Walker}, {Wester}, {Zhang}, \& {DES
  Collaboration}}]{BernardinelliCatalog}
---. 2020{\natexlab{b}}, \apjs, 247, 32, \dodoi{10.3847/1538-4365/ab6bd8}

\bibitem[{Brown(2001)}]{Brown2001}
Brown, M.~E. 2001, The Astronomical Journal, 121, 2804.
\newblock \url{http://stacks.iop.org/1538-3881/121/i=5/a=2804}

\bibitem[{Brown(2017)}]{Brown2017}
---. 2017, The Astronomical Journal, 154, \dodoi{10.3847/1538-3881/aa79f4}

\bibitem[{Brown \& Batygin(2019)}]{BB19}
Brown, M.~E., \& Batygin, K. 2019, The Astronomical Journal, 157, 62,
  \dodoi{10.3847/1538-3881/aaf051}

\bibitem[{{Dark Energy Survey Collaboration} {et~al.}(2016){Dark Energy Survey
  Collaboration}, {Abbott}, {Abdalla}, {Aleksi{\'c}}, {Allam}, {Amara},
  {Bacon}, {Balbinot}, {Banerji}, {Bechtol}, {Benoit-L{\'e}vy}, {Bernstein},
  {Bertin}, {Blazek}, {Bonnett}, {Bridle}, {Brooks}, {Brunner}, {Buckley-Geer},
  {Burke}, {Caminha}, {Capozzi}, {Carlsen}, {Carnero-Rosell}, {Carollo},
  {Carrasco-Kind}, {Carretero}, {Castander}, {Clerkin}, {Collett}, {Conselice},
  {Crocce}, {Cunha}, {D'Andrea}, {da Costa}, {Davis}, {Desai}, {Diehl},
  {Dietrich}, {Dodelson}, {Doel}, {Drlica-Wagner}, {Estrada}, {Etherington},
  {Evrard}, {Fabbri}, {Finley}, {Flaugher}, {Foley}, {Fosalba}, {Frieman},
  {Garc{\'{\i}}a-Bellido}, {Gaztanaga}, {Gerdes}, {Giannantonio}, {Goldstein},
  {Gruen}, {Gruendl}, {Guarnieri}, {Gutierrez}, {Hartley}, {Honscheid}, {Jain},
  {James}, {Jeltema}, {Jouvel}, {Kessler}, {King}, {Kirk}, {Kron}, {Kuehn},
  {Kuropatkin}, {Lahav}, {Li}, {Lima}, {Lin}, {Maia}, {Makler}, {Manera},
  {Maraston}, {Marshall}, {Martini}, {McMahon}, {Melchior}, {Merson}, {Miller},
  {Miquel}, {Mohr}, {Morice-Atkinson}, {Naidoo}, {Neilsen}, {Nichol}, {Nord},
  {Ogando}, {Ostrovski}, {Palmese}, {Papadopoulos}, {Peiris}, {Peoples},
  {Percival}, {Plazas}, {Reed}, {Refregier}, {Romer}, {Roodman}, {Ross},
  {Rozo}, {Rykoff}, {Sadeh}, {Sako}, {S{\'a}nchez}, {Sanchez}, {Santiago},
  {Scarpine}, {Schubnell}, {Sevilla-Noarbe}, {Sheldon}, {Smith}, {Smith},
  {Soares-Santos}, {Sobreira}, {Soumagnac}, {Suchyta}, {Sullivan}, {Swanson},
  {Tarle}, {Thaler}, {Thomas}, {Thomas}, {Tucker}, {Vieira}, {Vikram},
  {Walker}, {Wechsler}, {Weller}, {Wester}, {Whiteway}, {Wilcox}, {Yanny},
  {Zhang}, \& {Zuntz}}]{2016MNRAS.460.1270D}
{Dark Energy Survey Collaboration}, {Abbott}, T., {Abdalla}, F.~B., {et~al.}
  2016, \mnras, 460, 1270, \dodoi{10.1093/mnras/stw641}

\bibitem[{{DES Collaboration}(2005)}]{DES2005}
{DES Collaboration}. 2005, ArXiv e-prints.
\newblock \doarXiv{astro-ph/0510346}

\bibitem[{{Flaugher} {et~al.}(2015){Flaugher}, {Diehl}, {Honscheid}, {Abbott},
  {Alvarez}, {Angstadt}, {Annis}, {Antonik}, {Ballester}, {Beaufore},
  {Bernstein}, {Bernstein}, {Bigelow}, {Bonati}, {Boprie}, {Brooks},
  {Buckley-Geer}, {Campa}, {Cardiel-Sas}, {Castander}, {Castilla}, {Cease},
  {Cela-Ruiz}, {Chappa}, {Chi}, {Cooper}, {da Costa}, {Dede}, {Derylo},
  {DePoy}, {de Vicente}, {Doel}, {Drlica-Wagner}, {Eiting}, {Elliott}, {Emes},
  {Estrada}, {Fausti Neto}, {Finley}, {Flores}, {Frieman}, {Gerdes},
  {Gladders}, {Gregory}, {Gutierrez}, {Hao}, {Holland}, {Holm}, {Huffman},
  {Jackson}, {James}, {Jonas}, {Karcher}, {Karliner}, {Kent}, {Kessler},
  {Kozlovsky}, {Kron}, {Kubik}, {Kuehn}, {Kuhlmann}, {Kuk}, {Lahav}, {Lathrop},
  {Lee}, {Levi}, {Lewis}, {Li}, {Mandrichenko}, {Marshall}, {Martinez},
  {Merritt}, {Miquel}, {Mu{\~n}oz}, {Neilsen}, {Nichol}, {Nord}, {Ogando},
  {Olsen}, {Palaio}, {Patton}, {Peoples}, {Plazas}, {Rauch}, {Reil}, {Rheault},
  {Roe}, {Rogers}, {Roodman}, {Sanchez}, {Scarpine}, {Schindler}, {Schmidt},
  {Schmitt}, {Schubnell}, {Schultz}, {Schurter}, {Scott}, {Serrano}, {Shaw},
  {Smith}, {Soares-Santos}, {Stefanik}, {Stuermer}, {Suchyta}, {Sypniewski},
  {Tarle}, {Thaler}, {Tighe}, {Tran}, {Tucker}, {Walker}, {Wang}, {Watson},
  {Weaverdyck}, {Wester}, {Woods}, {Yanny}, \& {DES
  Collaboration}}]{2015AJ....150..150F}
{Flaugher}, B., {Diehl}, H.~T., {Honscheid}, K., {et~al.} 2015, \aj, 150, 150,
  \dodoi{10.1088/0004-6256/150/5/150}

\bibitem[{{Hamilton}({2019})}]{HamiltonPhD}
{Hamilton}, S. {2019}, PhD thesis, {University of Michigan}

\bibitem[{{Khain} {et~al.}(2020){Khain}, {Becker}, {Lin}, {Gerdes}, {Adams},
  {Bernardinelli}, {Bernstein}, {Franson}, {Markwardt}, {Hamilton}, {Napier},
  {Sako}, {Abbott}, {Avila}, {Bertin}, {Brooks}, {Buckley-Geer}, {Burke},
  {Carnero Rosell}, {Carrasco Kind}, {Carretero}, {Costa}, {Vicente}, {Desai},
  {Diehl}, {Doel}, {Flaugher}, {Frieman}, {Garc{\'\i}a-Bellido}, {Gaztanaga},
  {Gruen}, {Gruendl}, {Gschwend}, {Gutierrez}, {Hollowood}, {Honscheid},
  {James}, {Kuropatkin}, {Maia}, {Marshall}, {Menanteau}, {Miller}, {Miquel},
  {Plazas}, {Sanchez}, {Scarpine}, {Schubnell}, {Sevilla-Noarbe}, {Smith},
  {Sobreira}, {Suchyta}, {Swanson}, {Tarle}, {Walker}, {Wester}, \& {Dark
  Energy Survey Collaboration}}]{TaliClassification}
{Khain}, T., {Becker}, J.~C., {Lin}, H.~W., {et~al.} 2020, \aj, 159, 133,
  \dodoi{10.3847/1538-3881/ab7002}

\bibitem[{{Lawler} {et~al.}(2018){Lawler}, {Kavelaars}, {Alexandersen},
  {Bannister}, {Gladman}, {Petit}, \& {Shankman}}]{2018arXiv180200460L}
{Lawler}, S.~M., {Kavelaars}, J., {Alexandersen}, M., {et~al.} 2018, ArXiv
  e-prints.
\newblock \doarXiv{1802.00460}

\bibitem[{{Petit} {et~al.}(2011){Petit}, {Kavelaars}, {Gladman},
  {et~al.}}]{2011AJ....142..131P}
{Petit}, J.-M., {Kavelaars}, J.~J., {Gladman}, B.~J., {et~al.} 2011, \aj, 142,
  131, \dodoi{10.1088/0004-6256/142/4/131}

\bibitem[{Scholtz \& Unwin(2020)}]{PhysRevLett.125.051103}
Scholtz, J., \& Unwin, J. 2020, Phys. Rev. Lett., 125, 051103,
  \dodoi{10.1103/PhysRevLett.125.051103}

\bibitem[{{Schwamb} {et~al.}(2018){Schwamb}, {Jones}, {Chesley}, {Fitzsimmons},
  {Fraser}, {Holman}, {Hsieh}, {Ragozzine}, {Thomas}, {Trilling}, {Brown},
  {Bannister}, {Bodewits}, {de Val-Borro}, {Gerdes}, {Granvik}, {Kelley},
  {Knight}, {Seaman}, {Ye}, \& {Young}}]{LSSTsolarsystem}
{Schwamb}, M.~E., {Jones}, R.~L., {Chesley}, S.~R., {et~al.} 2018, arXiv
  e-prints, arXiv:1802.01783.
\newblock \doarXiv{1802.01783}

\bibitem[{{Shankman} {et~al.}(2017){Shankman}, {Kavelaars}, {Bannister},
  {Gladman}, {Lawler}, {Chen}, {Jakubik}, {Kaib}, {Alexandersen}, {Gwyn},
  {Petit}, \& {Volk}}]{OSSOS17}
{Shankman}, C., {Kavelaars}, J.~J., {Bannister}, M.~T., {et~al.} 2017, \aj,
  154, 50, \dodoi{10.3847/1538-3881/aa7aed}

\bibitem[{{Sheppard} \& {Trujillo}(2016)}]{ST16}
{Sheppard}, S.~S., \& {Trujillo}, C. 2016, \aj, 152, 221,
  \dodoi{10.3847/1538-3881/152/6/221}

\bibitem[{{Sheppard} {et~al.}(2016){Sheppard}, {Trujillo}, \&
  {Tholen}}]{2016ApJ...825L..13S}
{Sheppard}, S.~S., {Trujillo}, C., \& {Tholen}, D.~J. 2016, \apjl, 825, L13,
  \dodoi{10.3847/2041-8205/825/1/L13}

\bibitem[{Sheppard {et~al.}(2019)Sheppard, Trujillo, Tholen, \&
  Kaib}]{Sheppard_2019}
Sheppard, S.~S., Trujillo, C.~A., Tholen, D.~J., \& Kaib, N. 2019, The
  Astronomical Journal, 157, 139, \dodoi{10.3847/1538-3881/ab0895}

\bibitem[{{Trilling} {et~al.}(2019){Trilling}, {Gerdes}, {Trujillo},
  {Sheppard}, {Fuentes}, {Schlichting}, {McNeill}, {Juric}, {Holman}, {Lin},
  {Markwardt}, {Mommert}, {Oldroyd}, {Payne}, {Ragozzine}, {Rivkin}, \&
  {Schwamb}}]{DEEP2019}
{Trilling}, D., {Gerdes}, D., {Trujillo}, C., {et~al.} 2019, in EPSC-DPS Joint
  Meeting 2019, Vol. 2019, EPSC--DPS2019--395

\bibitem[{{Trujillo} \& {Sheppard}(2014)}]{st14}
{Trujillo}, C.~A., \& {Sheppard}, S.~S. 2014, \nat, 507, 471,
  \dodoi{10.1038/nature13156}

\end{thebibliography}

\newpage
\appendix

\begin{figure}[ht]
    \centering
    \includegraphics[width=0.8\textwidth]{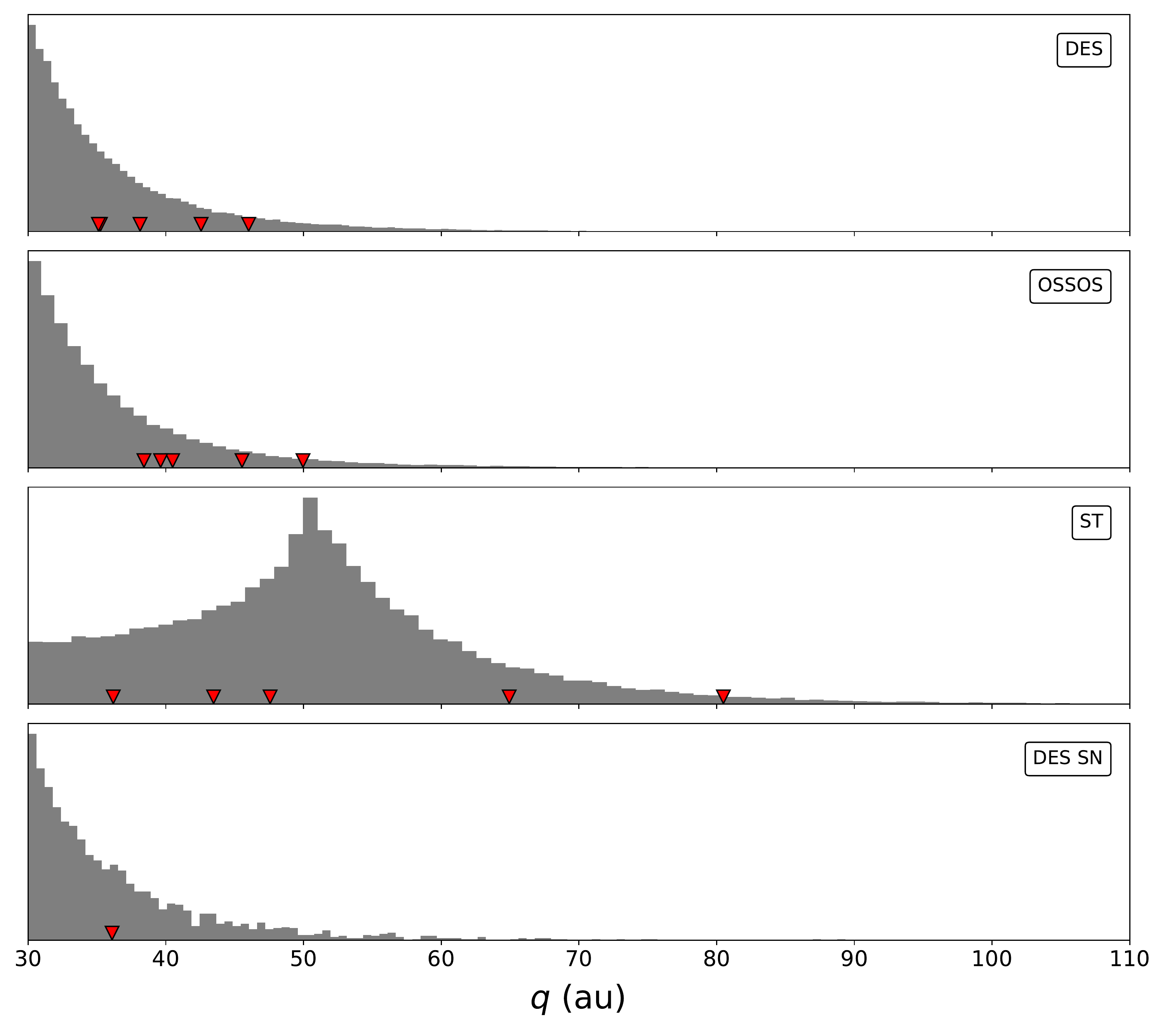} 
    \caption{Posterior pericenter distance distributions of simulated detections. The red triangles are the real ETNO detections by each survey. Note that DES has partially overlapping data points at $q\approx35$~au.}
    \label{fig:q_post}
\end{figure}

\begin{figure}[ht]
    \centering
    \includegraphics[width=0.8\textwidth]{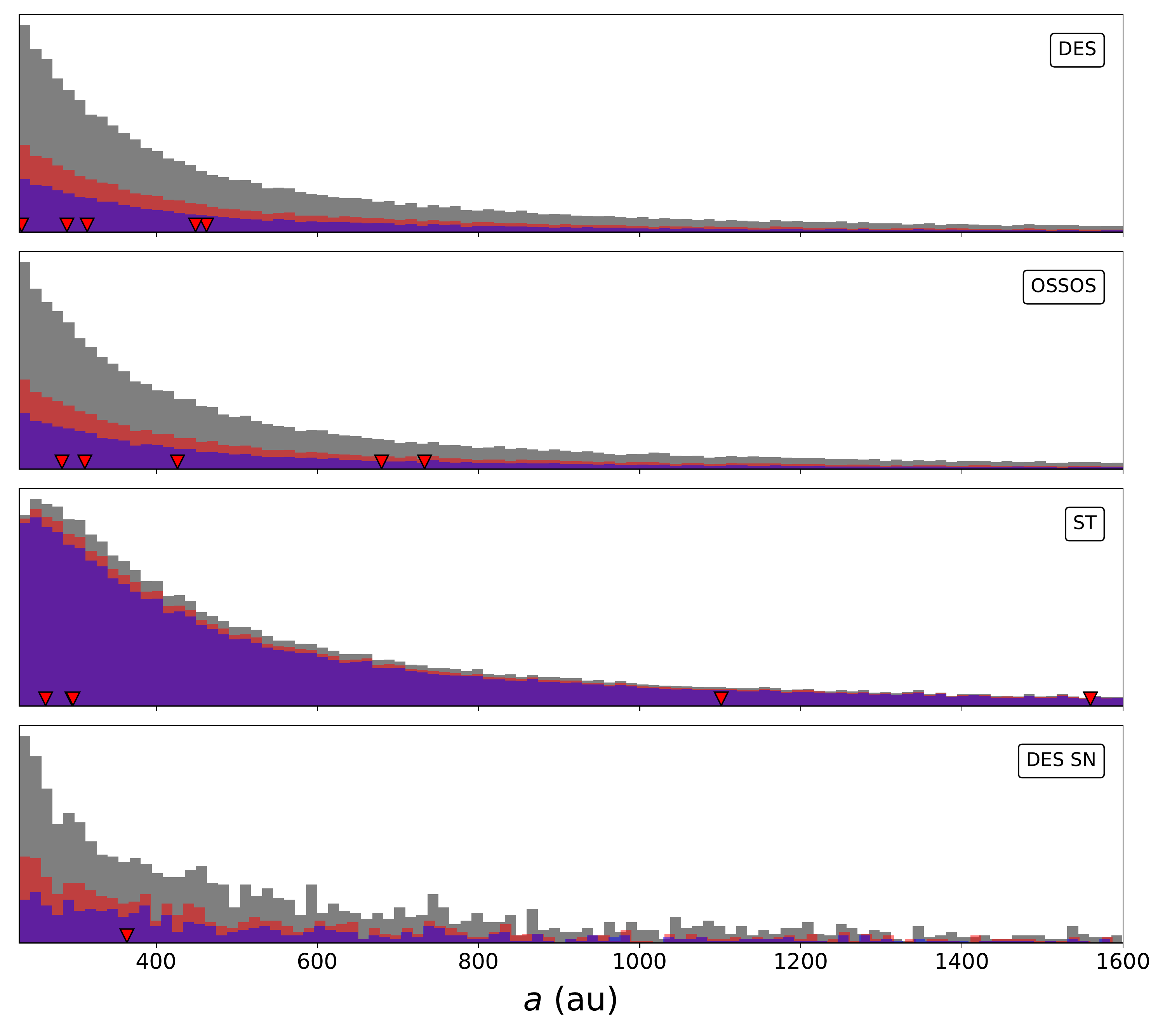} 
    \caption{Posterior semi-major axis distributions of simulated detections. The red triangles are the real ETNO detections by each survey. Note that ST has partially overlapping data points at $a\approx296$~au. The grey, red, and blue histograms correspond to cuts with $q > 30, 35$ and $38$ au, respectively.}
    \label{fig:a_post}
\end{figure}

\begin{figure}[ht]
    \centering
    \includegraphics[width=0.8\textwidth]{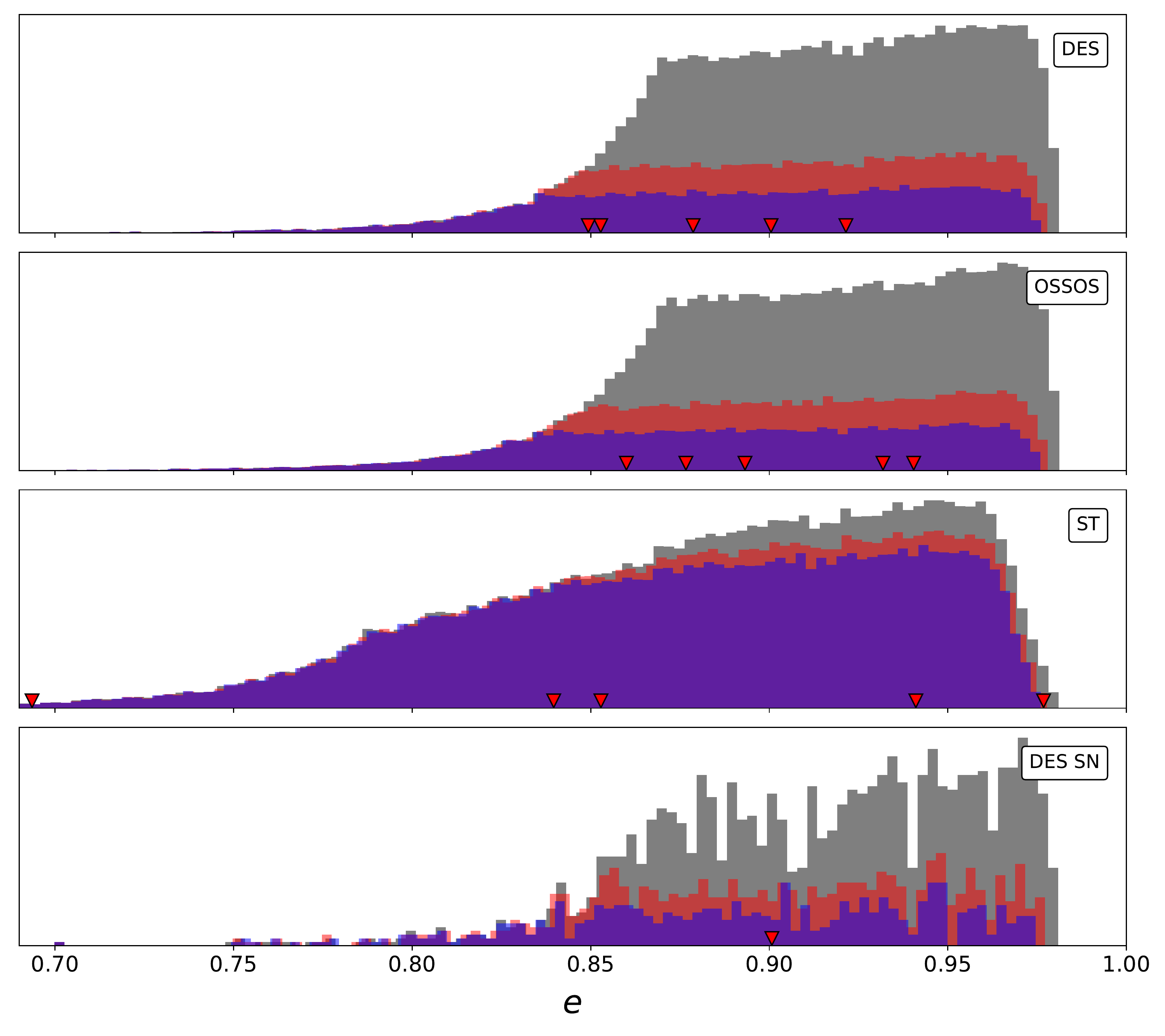} 
    \caption{Posterior eccentricity distributions of simulated detections. The red triangles are the real ETNO detections by each survey. Note that DES has overlapping data points at $e=0.85$. The grey, red, and blue histograms correspond to cuts with $q > 30, 35$ and $38$ au, respectively.}
    \label{fig:e_post}
\end{figure}

\begin{figure}[ht]
    \centering
    \includegraphics[width=0.8\textwidth]{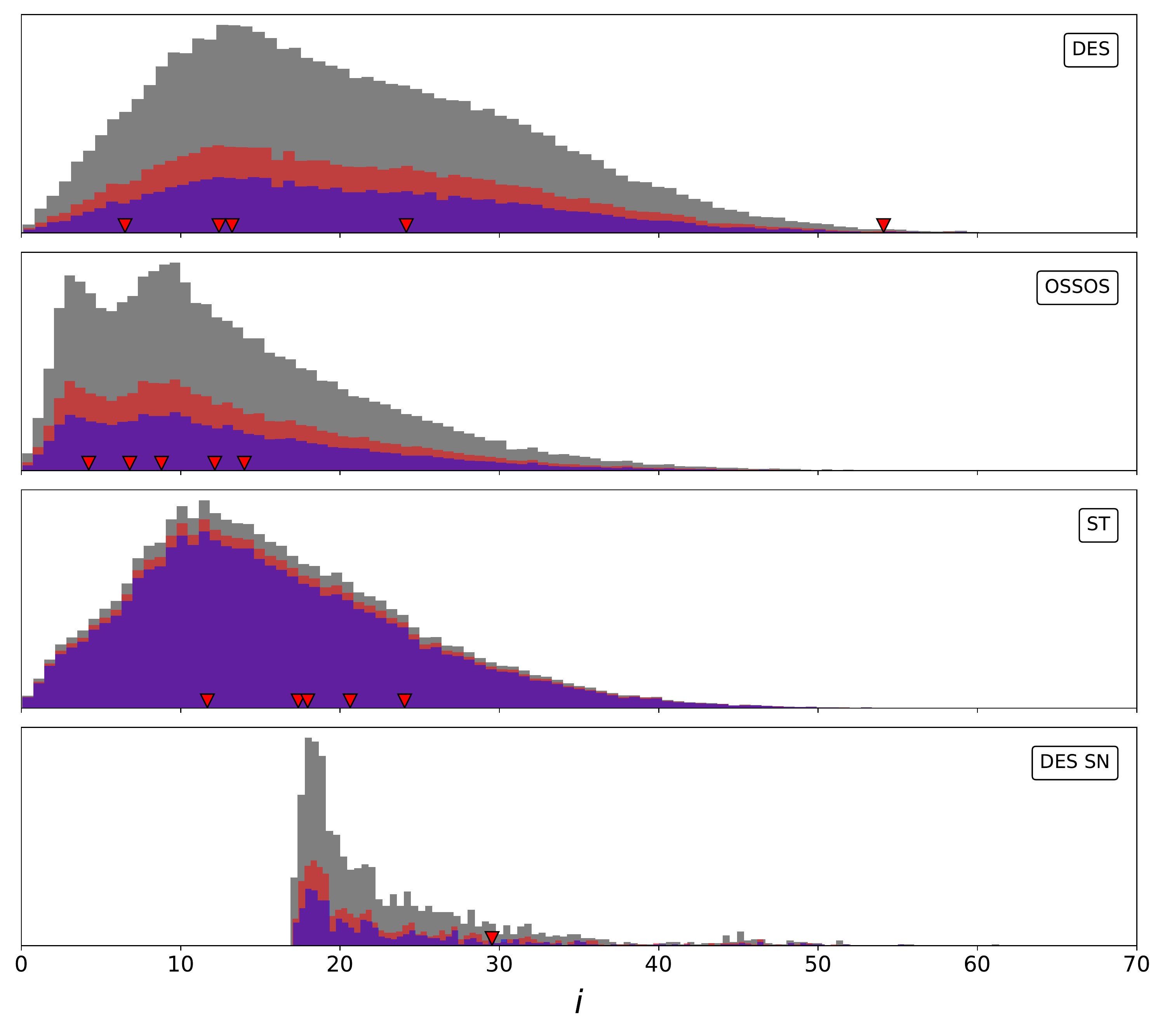} 
    \caption{Posterior inclination distributions of simulated detections. The red triangles represent the real ETNO detections by each survey. The grey, red, and blue histograms correspond to cuts with $q > 30, 35$ and $38$ au, respectively.}
    \label{fig:inc_post}
\end{figure}

\begin{figure}[ht]
    \centering
    \includegraphics[width=0.8\textwidth]{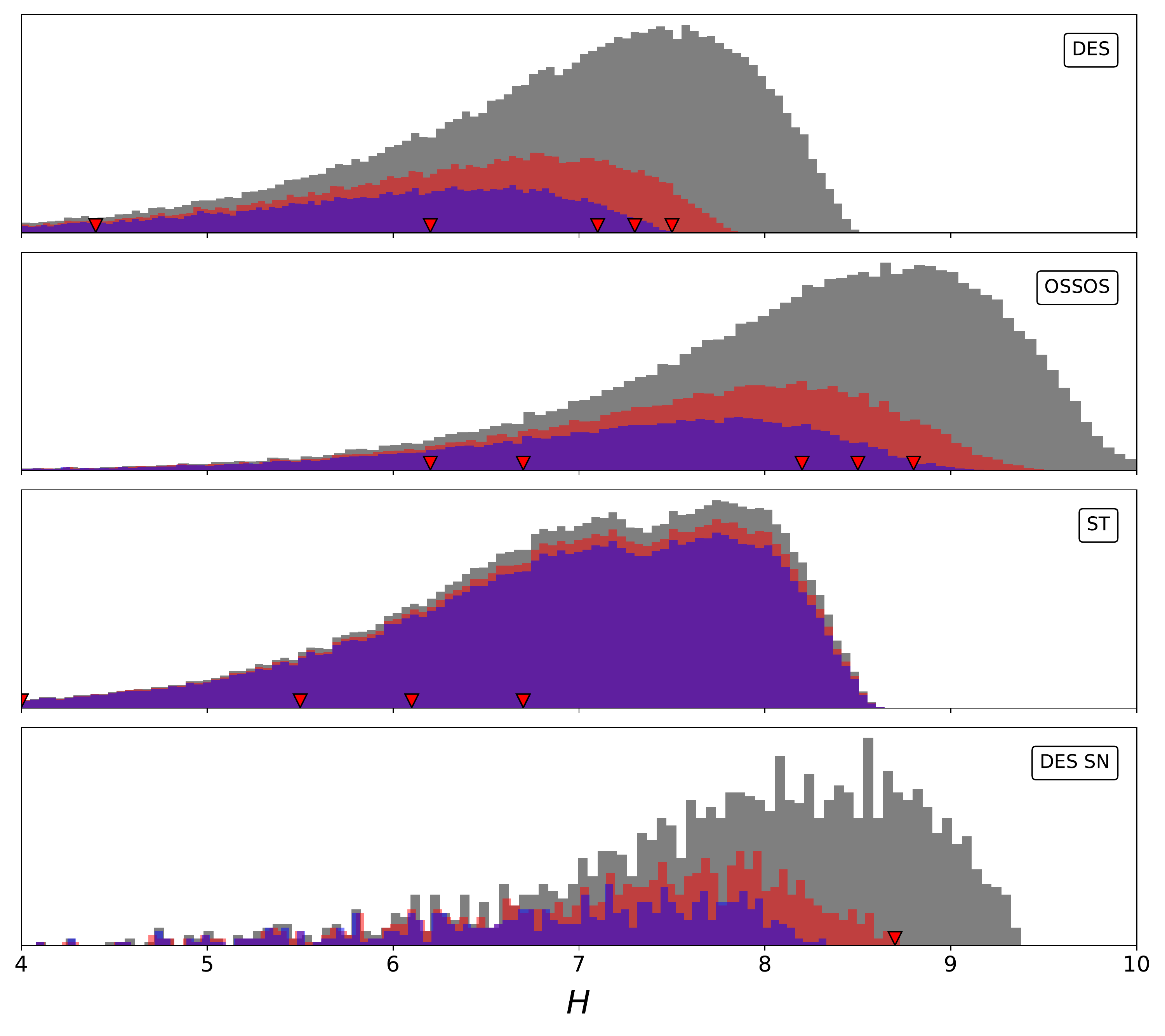} 
    \caption{Posterior absolute magnitude distributions of simulated detections. The red triangles represent the real ETNO detections by each survey. Note that ST has overlapping data points at $H=6.7$. The grey, red, and blue histograms correspond to cuts with $q > 30, 35$ and $38$ au, respectively.}
    \label{fig:H_post}
\end{figure}

\begin{figure}[ht]
    \centering
    \includegraphics[width=0.8\textwidth]{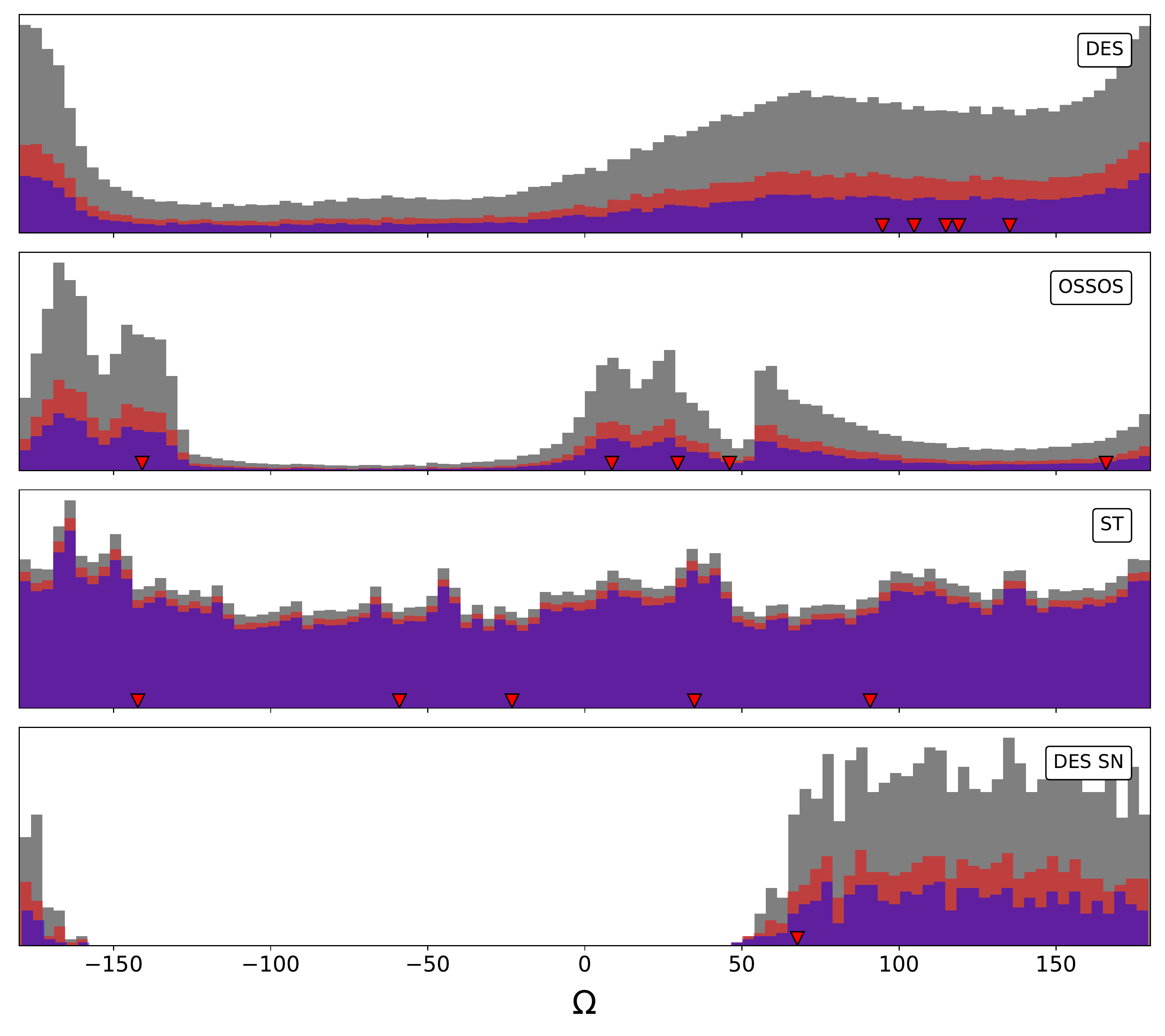} 
    \caption{Posterior longitude of ascending node distributions of simulated detections. The red triangles represent the real ETNO detections by each survey. The grey, red, and blue histograms correspond to cuts with $q > 30, 35$ and $38$ au, respectively.}
    \label{fig:node_post}
\end{figure}

\begin{figure}[ht]
    \centering
    \includegraphics[width=0.8\textwidth]{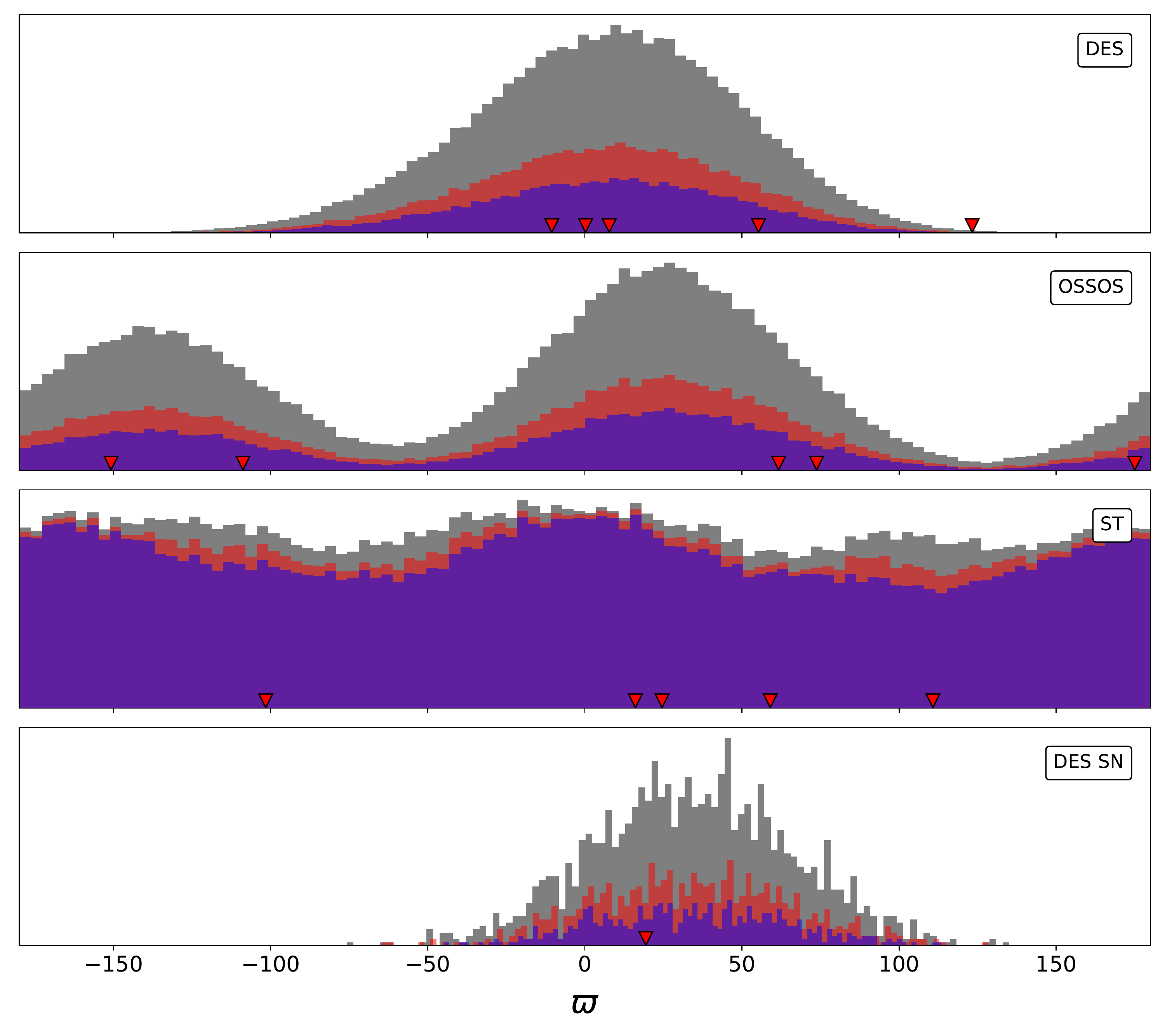} 
    \caption{Posterior longitude of pericenter distributions of simulated detections. The red triangles represent the real ETNO detections by each survey. The grey, red, and blue histograms correspond to cuts with $q > 30, 35$ and $38$ au, respectively.}
    \label{fig:varpi_post}
\end{figure}

\end{document}